\documentclass[11pt]{article}
\usepackage{graphicx}
\usepackage{amssymb}
\usepackage{amsmath}
\usepackage{graphicx}
\usepackage{amstext}
\usepackage{leftidx}
\usepackage{esint}
\usepackage[utf8]{inputenc}
\usepackage{color}
\usepackage{cite}

\numberwithin{equation}{section}

\renewcommand{\vec}[1]{{\bf #1}}       
\def\beq{\begin{eqnarray}}    
\def\eeq{\end{eqnarray}}      

\newcommand{\rL}{\rho_\Lambda}

\newcommand{\CC}{\Lambda}

\newcommand{\rv}{\rho_{\rm vac}}
\newcommand{\Pv}{P_{\rm vac}}
\newcommand{\rvo}{\rho^0_{\rm vac}}

\newcommand{\nueff}{\nu_{\rm eff}}

\newcommand{\bk}{{\bf k}}
\newcommand{\mpl}{m_{\rm Pl}}
\newcommand{\MPl}{{\cal M}_{\rm Pl}}

\newcommand{\be}{\begin{equation}}
\newcommand{\ee}{\end{equation}}

\newcommand{\EP}{V_{\rm eff}}
\newcommand{\tEP}{{V}_{\rm scal}}
\newcommand{\EPR}{V_{\rm eff}}

\newcommand{\new}[1]{{\textcolor{black}{#1}}}

\newcommand{\cH}{\mathcal{H}}

\newcommand{\txi}{\tilde{\xi}}

\newcommand{\astar}{a_{*}}

\newcommand{\rI}{\rho_I}

\newcommand{\ZPE}{V_{\rm ZPE}}
\newcommand{\CUV}{\Lambda_{\rm UV}}

\newcommand{\rV}{\rho_{\rm vac}}

\newcommand{\rVu}{\rho_{\rm vac}^{(1)}}

\usepackage{multido}
\makeatletter
\ams@newcommand{\vardot}[2]{%
  {\mathop{#2\kern0pt}\limits^{\vbox to-1.4\ex@{\kern-\tw@\ex@
   \hbox{\normalfont\multido{}{#1}{.}}\vss}}}}
\makeatother

\begin{document}



 \hyphenation{nu-cleo-syn-the-sis u-sing si-mu-la-te ma-king par-ti-cu-lar-ly
cos-mo-lo-gy know-led-ge e-vi-den-ce stu-dies be-ha-vi-or
res-pec-ti-ve-ly appro-xi-ma-te-ly gra-vi-ty sca-ling ha-ving theo-re-ti-cal
ge-ne-ra-li-zed re-gu-la-ri-za-tion mo-del mo-dels po-wers ex-cee-din-gly ho-we-ver me-tric pa-ra-me-ter  vacu-um per-for-ming appro-xi-ma-tion li-te-ra-tu-re pro-pa-ga-tor}




\begin{center}
{\bf \LARGE  The Cosmological Constant Problem and Running Vacuum in the Expanding Universe\footnote{ Invited review for Philosophical Transactions of the Royal Society  A, 2022,  to appear as a contribution  to  the theme issue  ``The Future of Mathematical Cosmology''.}} \vskip 2mm

 \vskip 8mm

\textbf{Joan Sol\`a Peracaula}

\vskip 0.5cm
Departament de F\'isica Qu\`antica i Astrof\'isica, \\
and \\ Institute of Cosmos Sciences,\\ Universitat de Barcelona, \\
Av. Diagonal 647, E-08028 Barcelona, Catalonia, Spain

\vskip0.5cm

\vskip0.4cm

E-mail:  sola@fqa.ub.edu

 \vskip2mm

\end{center}
\vskip 15mm

\begin{quotation}
\noindent {\large\it \underline{Abstract}}.
It is well-known that quantum field theory (QFT) induces a huge value of the cosmological constant, $\Lambda$, which is outrageously inconsistent with cosmological observations. We review here some aspects of this fundamental theoretical conundrum (`the cosmological constant problem') and strongly argue in favor of the possibility that the cosmic vacuum density $\rho_{\rm vac}$ may be mildly evolving  with the expansion rate $H$.  Such a  `running vacuum model' (RVM) proposal predicts an effective dynamical dark energy without postulating new ad hoc fields (quintessence and the like).  Using the method of adiabatic renormalization within QFT in curved spacetime we find  that  $\rho_{\rm vac}(H)$ acquires a  dynamical component ${\cal O}(H^2)$  caused by the quantum matter effects. There are also  ${\cal O}(H^n)$ ($n=4,6,..$)  contributions,  some of which may trigger inflation  in the early universe.  Remarkably, the evolution of the adiabatically renormalized  $\rho_{\rm vac}(H)$ is not affected by dangerous terms proportional to the quartic power of the masses ($\sim m^4$) of the fields.  Traditionally, these terms have been the main source of trouble as they are responsible  for the extreme fine tuning feature of the  cosmological constant problem.   In the context under study,  however,  the late time $\rho_{\rm vac}(H)$ around $H_0$ is given by a dominant term ($\rho_{\rm vac}^0$)  plus the aforementioned  mild dynamical component  $\propto \nu (H^2-H_0^2)$  (with $|\nu|\ll1$), which makes the RVM to mimic quintessence. Finally,  on  the phenomenological side we show that the RVM may be instrumental in alleviating some of the most challenging problems (so-called `tensions') afflicting nowadays the observational consistency of the `concordance'  $\Lambda$CDM model, such as the $H_0$ and $\sigma_8$ tensions.

\end{quotation}
\vskip 5mm

\newpage

\tableofcontents

\newpage


\section{Introduction}

The standard or concordance model of cosmology, or $\CC$CDM model,  is a rather  successful  theoretical framework for the phenomenological description of the Universe\,\cite{Peebles1984,Peebles1993,KraussTurner1995,OstrikerSteinhardt1995}.  The current paradigm is almost forty years old, but it became experimentally solid only in the mid nineties\,\cite{Turner2021}. Being accepted by most cosmologists, it is also called the `concordance' cosmological model. Crucial ingredients of it, however, lack of observational evidence and/or of a proper theoretical understanding  based on fundamental principles.  Such is the situation with dark matter, of  which we still have no direct evidence. No less preoccupying is the theoretical status of  the cosmological constant (CC) term, $\CC$,  in Einstein's equations.  Despite it is permitted by general covariance, nothing is firmly known about its possible physical origin.  The roots of the problem reside in its interpretation  as a quantity which is connected with the vacuum energy density (VED), $\rv$, a fundamental concept in  Quantum Field Theory (QFT).  In fact, the  notion of vacuum energy in cosmology is a most subtle concept which  has challenged theoretical physicists and cosmologists for more than half of a century, specially with the advent of Quantum Theory and in general with the development of the more sophisticated conceptual machinery of  QFT.    The proposed connection is $\rv=\CC/(8\pi G_N)$,  where $G_N$ is Newton's constant.  Consistent measurements of this parameter made independently in the last twenty years using distant type Ia supernovae (SnIa)\,\cite{SNIa}  and the anisotropies of the cosmic microwave background (CMB) \cite{PlanckCollab}, have put the foundations of the concordance  $\CC$CDM model of cosmology\cite{Peebles1993}.  The model  is continuously being tested and must be  confronted  with the  hints of new physics that may come from the many cosmic messengers\,\cite{CostAction}.

Historically, the $\CC$-term in the gravitational  field equations was introduced by A. Einstein 105 years ago\,\cite{Einstein1917}, but the  ``cosmological constant problem'' (CCP) as such was first formulated 50 years later by Y. B. Zeldovich\,\cite{Zeldovich1967a,Zeldovich1967b}.  The CCP is the baffling  realization that the successful QFT methods applied to the world of the elementary particles seem to predict an effective value for $\rv$ which is devastatingly much larger than the value of the current critical density of the universe (which the  vacuum energy density should be comparable with) \cite{Weinberg89,Witten2000,Sahni2000,PeeblesRatra2003,Padmanabhan2003,Copeland2006,JMartin2012,JSPRev2013}.   The first theoretical models trying to cope with the CCP from a   field theoretical viewpoint were models of  dynamical scalar fields, see particularly the prototype model by Dolgov in\,\cite{Dolgov82} and other attempts which appeared shortly afterwards\,\cite{Abbott85,Banks85, PSW87,Barr87,Ford87,Sola89,Sola90}. They endeavored to explain  the value of  the vacuum energy in terms of a dynamical mechanism, e.g. an effective potential of a cosmic scalar field which settles down automatically (hence without fine tuning) to the present small value, $\rvo$.   As an example, see  the ``cosmon'' attempt in \cite{PSW87} and its ulterior discussion by Weinberg in \cite{Weinberg89}.   Unfortunately, no model of this kind ever succeeded as yet in achieving such an unattainable dream, as in fact an unsurmountable stumbling block was placed in their way in the form of a ``no-go theorem''  formulated by Weinberg\,\cite{Weinberg89}.

\new{In the meantime a new generation of scalar field models gave rise to the notion of quintessence. They  appeared in the market in two different waves separated by almost a  decade\,\cite{Wetterich88,RatraPeebles88,Frieman95,Caldwell98}.  Quintessence models were all aimed at  a much more modest purpose: rather than trying to explain the value of $\rvo$ (the hardest aspect of the CCP, i.e. the  ``old cosmological constant problem'') they  focused instead on the so-called  ``cosmic coincidence problem''\cite{Steinhardt1997,Steinhardt2003}, viz. why $\rvo$ is so close to the present matter density.  It is, however,  disputable whether this is a problem at all,  see e.g. \cite{Ishak2005,Lineweaver2007}.  In any case, phenomenological solutions to the purported problem have been put forward from very different perspectives, see e.g. \cite{Zlatev1999,Steinhardt99,Zimdahl2001,LXCDM2006}, among many others.  Let us also mention that in the interreign period mediating in between the appearance of the first scalar field  models aiming at solving the CCP dynamically and the next delivery of scalar field proposals trying to explain the coincidence problem, a very large family of  phenomenological  `$\CC(t)$-models'  were suggested,  sometimes simultaneously to quintessence and the like.  In those models, the  $\CC$ term is not replaced by a scalar field but it is assumed to acquire some \textit{ad hoc} time dependence (sometimes through the scale factor and/or the Hubble rate, or just directly in terms of the cosmic time).  The  `$\CC(t)$-class'  of models is often referred to as  `Decaying Vacuum Cosmologies', see e.g.\,\cite{Fujii82,OzerTaha87,Freese87,ChenWu90,Carvalho92}. The idea of a time evolving cosmological term  is actually much older and traces back to Bronstein in 1933\,\cite{Bronstein33}. For a review of many of these models, see e.g.  \cite{Overduin98}  and references therein.  The $\CC(t)$-class  has no obvious relation with fundamental physics, but it was useful  to  illustrate the possible effects implied by a time-evolving cosmological term.  These models, however,  should not be confused with the ``running vacuum models'' (RVMs), in which the  dynamical dependence of $\CC$, or more precisely of the VED, stems from explicit computation of quantum effects  in QFT in curved spacetime\,\cite{CristianJoan2020, CristianJoan2022}.  Therefore, despite some existing  confusion in the literature, the RVMs are to be understood in that (much more) restricted sense.  In point of fact,  only the RVMs will be dealt with in the current article, see \,\cite{JSPRev2013} for an introduction.}

In the years after the discovery of the accelerated expansion of the universe a flood of proposals invaded hastily the cosmology market: it  certified the birth of the miscellaneous notion of `dark energy' (DE),  see \cite{DEBook} and the long list of references therein. Its purpose was to replace (in `infinitely' many different disguises) the function made by the $\CC$-term in Einstein's equations.  Because of the CCP, the vacuum energy option for the DE became  outcast as if being blamed of all evils, particularly of the acute fine tuning problem. However, the criticisms usually had nothing better to offer, except to defend tooth and nail some particular form of DE without  improving an inch the fine-tuning illness -- which for some mysterious reason was (exclusively) ascribed to the CC option.  But this is unfair, of course. All devised  forms of DE are, in principle, plagued with the same fine tuning  illness and to a degree which is no lesser than the CC itself.  Paradoxically enough,  it turns out that the vacuum energy can actually escape such a fate, as we will show here,  after it is properly renormalized\,\cite{CristianJoan2020, CristianJoan2022}. The vacuum is in fact the most fundamental notion of QFT. Thus,  a correct description of the CCP using the quantum vacuum and the machinery of QFT should be the most desirable approach  to tackle the big CC conundrum  from first principles.

One expects that any sensible proposal should largely preserve the  concordance $\CC$CDM model, which  is formulated in  the context of the Friedman-Lemaitre-Robertson-Walker (FLRW) framework and is deeply ingrained in the General Relativity (GR) paradigm.  However, one of the most important drawbacks of GR  is that it is a non-renormalizable theory.  This can be considered a  theoretical impediment for GR to be categorized as a fundamental theory of gravity and hence it adversely impacts on the status of $\CC$CDM as well.  GR cannot properly describe the short distance effects of gravity (the ultraviolet regime, UV), only the large distance effects (or infrared regime, IR).  As a consequence,  GR  alone cannot provide a framework for quantizing gravity (the metric tensor field of spacetime)  together with the rest of the fundamental  interactions  (assuming of course that gravity can be  quantized on conventional grounds). Nonetheless a first (rougher, albeit effective) approach is to treat gravity as a classical (external or background) field and quantize  the matter fields only. This is,  after all, the essence of the time-honoured  semiclassical approach called  QFT in curved spacetime, see e.g. \cite{BirrellDavies82,ParkerToms09,Fulling89,MukhanovWinitzki07,ParkerCargese1978,Ford1997} for a review.

It will be our approach here too. Notwithstanding its limitations, we will still be able to estimate the quantum effects from the matter fields and their contribution to the VED, $\rv$, and  will enable us to renormalize these effects (originally UV-divergent)  and obtain  finite quantities that can be better compared with observations.  Even more, the renormalization group (RG) flow driving the VED  in our framework  will be seen to be free from the troublesome $\sim m^4$ contributions (viz. proportional to the quartic powers of the masses of the quantized matter fields) which would otherwise be obtained e.g. using the Minimal Subtraction (MS)  scheme and other regulators.  This was first shown in the literature in Ref.\,\cite{CristianJoan2020} and further extended in\,\cite{CristianJoan2022}.   \new{In these works, the VED appears as an expansion in powers of $H$ and its time derivatives: $\rv=\rv(H, \dot{H}, \ddot{H},...)$.  The expansion terms are  the result of quantum effects computed within the context of QFT in curved spacetime.  For the present universe,  $\rv$ greatly simplifies and it   appears   as the sum of a  dominant term plus a dynamical component which varies as $\sim \nu H^2\mpl^2$, where $\nu$  is a small (dimensionless) and computable coefficient (playing the role of $\beta$-function coefficient of the VED running) and  $\mpl$ is the usual Planck mass ($\mpl=G_N^{-1/2}$) in natural units.  Such a result perfectly fulfils the above condition of being  compatible to a great extent  with the $\CC$CDM, as  it entails only a small departure from the main CC term.  No less remarkable is the fact that the quantum vacuum is predicted to be mildly dynamical, and such dynamics can be effectively evaluated from first principles since the parameter  $\nu$ can be accounted for using QFT in curved spacetime.}

As long as the QFT vacuum approach proposed here turns out to be free from the awkward $\sim m^4$ effects that are obtained in oversimplified renormalization treatments which prove inappropriate for the CCP, the renormalized result which ensues appears theoretically distinct and  tantalizing.  Remember that the CCP\cite{Weinberg89}  is one of the hardest and longstanding mysteries of theoretical physics,  and proves to be a  most serious obstruction to reconcile theoretical cosmology with particle physics\,\cite{JSPRev2013}.  Somehow one of the aspirations of the presaged  `third cosmological paradigm' upcoming in the near future\,\cite{Turner2021} ought to be to bring some light  on  improving such a fundamental and longstanding thorny issue of  theoretical cosmology.

The more tame  QFT formulation of the quantum vacuum that will be attained here is not only highly convenient from the theoretical  standpoint, but also from  the observational side. It has long been known that  there appear to exist discrepancies  or ``tensions'' between the results of data analyses from Planck Collaboration, based on $\Lambda$CDM~\cite{PlanckCollab}, and the local direct measurements of the Hubble parameter today, the so called $H_0$ tension. Not only so, there exist  also tensions in the large scale structure (LSS) growth data, the so-called $\sigma_8$ tension. For reviews of these tensions, see~\cite{TensionsLCDM,Intertwined8,IntertwinedH0,PerivoSkara2021,ValentinoReviewTensions,TensionsJSP2018,WhitePaper2022}, for example.   Although such discordancies in the arena of the standard model might eventually receive more mundane astrophysical explanations~\cite{efst,Freedman}, or even disappear into thin air when in the near future more  data will be available, it has been admitted already that they may well be passing the point of being attributable to a fluke\,\cite{Riess2019}.   Intriguingly enough,  the RVM can help smoothing out these tensions and improving the fits to the cosmological data\,\cite{EPL2021,CQG2020,MNRAS2018,EPL2018,PLB2017,ApJL2016,RVM2015,RVMphenoOlder,Polarski2012}.

The upshot is that the quantum vacuum proposal, which we will describe in what follows,  is theoretically appealing and  has the capacity to impinge positively on the description of the modern cosmological data.  Such a framework is precisely the aforementioned running vacuum model of cosmology (RVM), see\,\cite{JSPRev2013} for an ample introduction with many references,  and \,\cite{JSPRev2014,JSPRev2015,JSPRev2016} for  expositions with an emphasis on phenomenological applications.
The state-of-the-art-phenomenological performance of the RVM (and related models mimicking its behavior) in the last few years can be appraised in Ref.\,\cite{EPL2021} and can be complemented with the previously mentioned works. The model has demonstrated a sustained ability to compete with the concordance $\CC$CDM model.   On the theoretical side, the  RVM  was originally motivated by semi-qualitative RG arguments~\cite{ShapSol}, which were later on further elaborated  within an action functional approach\,\cite{Fossil08}. However, the theoretical underpinning of the RVM within the full-fledged context of  QFT in curved spacetime appeared only very recently, see \cite{CristianJoan2020} and also the very comprehensive exposition \cite{CristianJoan2022}, where the topics about the renormalization of the vacuum energy in cosmological spacetime and the implications on the cosmological constant problem are taken up in extenso and  in depth. For related studies about renormalization in curved spacetime, see e.g. \cite{Babic2005,Maggiore2011,Bilic2011,KohriMatsui2017,Antipin2017,FerreiroNavarroSalas}.
In this summarized account, I will expose the emergence of the  RVM from the mentioned  QFT calculations  in  FLRW spacetime, along with the most recent phenomenological applications. Before facing the major difficulties to be encountered in curved spaces, I will first review  some basic aspects related to the vacuum energy in flat (Minkowskian) spacetime.

The structure of this presentation is as follows. In section \ref{Sec:FlatSpaceAnalysis} we discuss the vacuum energy in flat spacetime. In section \ref{sec:EMT} we study the non-minimally coupled  scalar field in FLRW spacetime. The renormalization of the zero-point energy in curved spacetime and the connection with the RVM are described in sections \ref{eq:RegZPE}  and \,\ref{sec:RenormEMT}, respectively.  In Sec.\,\ref{Sec:EffectiveActionl} we retake the renormalization from the perspective  of the effective action formalism and confirm the consistence between the two procedures.  The novel feature of  RVM-inflation in QFT is addressed in section \ref{sec:Inflation}. Subsequently, in  section \ref{sec:PhenoApplications}, we put  the  RVM to the test and  show that it can alleviate the $\sigma_8$ and $H_0$ tensions.    Finally, section \ref{sec:conclusions} contains our conclusions and outlook.

\new{{\bf Conventions}: We use the same geometric conventions  as in\,\cite{CristianJoan2020,CristianJoan2022}. They correspond to  $(+, +, +)$ in the standard classification by Misner-Thorn-Wheeler\,\cite{MTW}. Thus, the signature of the metric  $g_{\mu\nu}$ is  $(-, +,+,+ )$ ,  Riemann  tensor
$R^\lambda_{\,\,\,\,\mu \nu \sigma} = \partial_\nu \, \Gamma^\lambda_{\,\,\mu\sigma} + \Gamma^\rho_{\,\, \mu\sigma} \, \Gamma^\lambda_{\,\, \rho\nu} - (\nu \leftrightarrow \sigma)$;  Ricci tensor, $R_{\mu\nu} = R^\lambda_{\,\,\,\,\mu \lambda \nu}$; and  Ricci scalar,  $R = g^{\mu\nu} R_{\mu\nu}$.   Spatially flat three-dimensional geometry will be asumed throughout.
Natural units ($\hbar=c=1$) will  be assumed as well, except when emphasis is made on the  nature of some quantities (such as e.g. in the next section).}

\section{Vacuum energy in flat spacetime: ZPE and SSB}\label{Sec:FlatSpaceAnalysis}

The CCP is a difficult problem, it is connected with the notion of  vacuum energy, but it actually goes beyond it because it involves gravity as well.  Let us flesh out the basic ingredients of the problem in QFT in flat, Minkowskian, spacetime first.  We shall commence reviewing the \textit{zero-point energy} (ZPE)  for a free field, and afterwards we will discuss the contribution to the vacuum energy density (VED) from a scalar field under \textit{spontaneous symmetry breaking} (SSB).   Curved space considerations will be dealt with in the next sections.  Our first observation is that the  ZPE is always present, whereas the SSB may or may not contribute depending on whether the Brout-Englert-Higgs (BEH) mechanism\,\cite{BEH-Mechanism64}  is active or not  to trigger the spontaneous breaking of a gauge symmetry. Let us assume for the moment that both contributions are present.   In QFT the field
$\phi$ becomes a quantum field operator $\hat{\phi}$ and
therefore its ground state value must now be interpreted
as the vacuum expectation value (VEV){:}
$<\phi>\equiv\langle 0|\hat{\phi}|0\rangle$. However,
the theory can still be handled as if $\phi$ were a classical (mean)
field provided its classical potential  is renormalized into the
effective potential:  $V\rightarrow
 \EP=V+\hbar\,V_1+ \hbar^2\,V_2+ \hbar^3\,V_3+...$
The quantum effects to all orders of perturbation theory arrange
themselves in the form of a loopwise expansion where the number of
loops is tracked by the powers of $\hbar$. Thus, at one loop we have
only one power of $\hbar$, at two loops we have two powers of
$\hbar$ etc. For $\hbar=0$, however, there are no loops and the
effective potential just reduces to the classical potential, $V$,
given e.g. by the standard Higgs potential (associated to the BEH mechanism)  in the electroweak standard model,  namely $V(\phi)=\frac12\,m^2\,\phi^2+\frac{1}{4!}\,\lambda\,\phi^4 \ (\lambda>0)$. Notice that  $m^2<0$, if there is SSB. In that case the physical mass squared  is $m_H^2=-2m^2>0$ (see e.g. \cite{JSPRev2013} for a review).
On the other hand, the loop terms in $\EP$ can be split into
two independent contributions, to wit: one having loops with no external
legs, {i.e. the vacuum-to-vacuum diagrams  $V_P^{(i)}$, which depend on a collection $P$ of parameters but not on $\phi$. This part is the ZPE. The other part carries
loops with external legs of the matter field $\phi$, i.e.\ the loop corrections $V_\text{scal}^{(i)}(\phi)$ to the
classical potential. Therefore, for each loop contribution we can write
\begin{equation}\label{splitloop}
V_1=V_P^{(1)}+V_\text{scal}^{(1)}(\phi)\,,\ \ \ \
V_2=V_P^{(2)}+V_\text{scal}^{(2)}(\phi)\,,\ \ \ \
V_3=V_P^{(3)}+V_\text{scal}^{(3)}(\phi)...\,.
\end{equation}
As a result, the effective potential $\EP$ at the quantum
level splits naturally into two parts, one which is
$\phi$-independent and another that is $\phi$-dependent:
\begin{equation}\label{EPZPE}
\EP(\phi)=\ZPE+\tEP(\phi)\,,
\end{equation}
in which  $\ZPE=\hbar\,V_P^{(1)}+\hbar^2\,V_P^{(2)}+\hbar^3\,V_P^{(3)}+....$ is the  aforementioned ZPE contribution, which consists of the
`bubble'-type (vacuum-to-vacuum) diagrams to all orders in perturbation theory.  Moreover,  $\tEP(\phi) = V(\phi)+\hbar\,V_{\rm scal}^{(1)}(\phi)+\hbar^2\,V_{\rm scal}^{(2)}(\phi)+ \hbar^3\,V_{\rm
scal}^{(3)}(\phi)+...$.  If there is no SSB ($m^2>0$ in the Higgs potential),  the ground state is at $\phi=0$ and hence $\tEP(\phi)=0$ (as all these diagrams carry external legs).  In this case the VED  is given by  $\EP(\phi=0)=\ZPE$.

The ZPE is  a pure quantum effect: it vanishes if there is no quantum theory,
$\hbar=0$. It is the most genuine quantum vacuum contribution.  The result can only
depend on the set of parameters $P$ of the theory (such as masses and coupling constants), but
not on $\phi$ itself since it enters virtually the loop propagators. In Minkowskian spacetime the one-loop effect $V_P^{(1)}$ in \eqref{splitloop} is
obtained by ignoring classical gravity since no external tails of the metric can be attached in this case to the bubble diagram
(in contrast to a curved background).  If we move to the continuum and regularize  the infinite
sum $(1/2)\sum_{\bf k} \hbar\omega_{\bf k}=(1/2)\sum_{\bf k} \hbar
\sqrt{{\bf k}^2+m^2}$ by means of a  UVcutoff ${\CUV}$  (which should not be confused with the CC term $\CC$), we can trivially integrate the solid angle with the result
\begin{eqnarray}\label{eq:ZPE1loop}
 V_P^{(1)} &=& \frac{1}{2}\int \frac{{\rm d}^3k}{(2\pi)^3}\, \sqrt{{\bf
k}^2+m^2}=\frac{1}{4\pi ^2}\int _0^{\CUV} {\rm d}k\, k^2
\sqrt{k^2+m^2}\nonumber\\
& = &\frac{\CUV^4}{16
\pi^2}\left(1+\frac{m^2}{\CUV^2}-\frac14\,\frac{m^4}{\CUV^4}\,\ln\frac{\CUV^2}{m^2}
+\cdots \right)\,,
\end{eqnarray}
where in the expansion in powers of $m/\CUV$ we have explicitly kept the
important term $\sim m^4\ln m^2$ because this one does not depend on any
power of the cutoff and therefore it will remain upon
removing the cutoff or any sensible regulator within a renormalization scheme.  The above expression is the
cutoff-regularized result for the Minkowskian  ZPE at one loop.
Although any scheme is in principle valid, not all of them have the
property that the renormalized quantities are in good correspondence with
the physical quantities in the particular framework of the calculation.
For example, it makes no much sense to renormalize  QCD (the gauge theory
of strong interactions) in the on-shell scheme because we never find
quarks and gluons on mass shell!  One is forced to use an off-shell
renormalization scheme. This may complicate the physical interpretation,
for there can  be a nontrivial gap between the renormalized
parameters and the physical ones.  This problem is particularly acute in cosmology,  where gravity should
play a role.  Taken at face value, the result (\ref{eq:ZPE1loop}) is far from having any obvious physical meaning, not even in flat spacetime.
 It needs  proper regularization (a rough cut-off, for instance,  breaks Lorentz invariance) and also proper  renormalization since it diverges for $\CUV\to\infty$. If that is not enough,
it badly needs  a sensible generalization in the context of curved spacetime.  The result \eqref{eq:ZPE1loop} says nothing at all about the cosmological constant and the vacuum energy in the expanding universe.

Even though we could preserve Lorentz invariance using an `educated cut-off' by regularizing \`a la Pauli-Villars, let us adopt the MS scheme (see e.g.\,\cite{Collins84,Sterman93,Brown92})  with dimensional regularization (DR), as it also preserves Lorentz invariance and leads to the same result.  The one-loop calculation of the ZPE using DR  in $n$ spacetime dimensions renders
\begin{equation}\label{Vacfree2}
\hbar\, V_P^{(1)}=
\frac12\,\mu^{4-n}\, \int\frac{d^{n-1} k}{(2\pi)^{n-1}}\,
   \hbar\,\sqrt{{\bf k}^2+m^2}
= \frac12\,\beta_{\rL}^{(1)}\,\left(-\frac{2}{4-n}
-\ln\frac{4\pi\mu^2}{m^2}+\gamma_E-\frac32\right) \,,
\end{equation}
where $\gamma_E$ is Euler's constant, and
$\beta_{\rL}^{(1)}=\frac{\hbar \,m^4}{2\,(4\pi)^2}$
is the one-loop coefficient of the $\beta$-function for the parameter $\rL$ (see below).  The above  (flat spacetime) result is well-known\,\cite{ShapSol,Akhmedov2002,Sirlin2003}; in actual fact, it is collected in textbooks  since long\,\cite{Brown92}.  In it,  $\mu$ is the arbitrary  't Hooft's mass unit of dimensional
regularization\,\cite{Collins84}, and $n\rightarrow 4$ is
understood in the final result. Obviously, the ZPE is UV-divergent in that limit, as there is a pole at $n=4$.  Renormalization is thereby imperative.

A more formal  method of derivation (which eases its generalization to QFT in curved spacetime, cf. Sec.\,\ref{Sec:EffectiveActionl})  can be achieved starting from the effective action $W$ of the theory\,\cite{Coleman,BooksQFT} computed within the effective potential approach, i.e.  we approximate $W=\int d^4x (-\hbar V_1)= \Omega  (-\hbar V_1)$ , where  $\Omega$ the spacetime volume  and
\begin{equation}\label{eq:V1}
V_1=-\,\frac{i}{2}\,\Omega^{-1}\,Tr\ln{\cal K}=-\,\frac{i}{2}\,\Omega^{-1}\,\int d^n x \lim_{x\to x'}\,\ln [{\cal K}(x,x')]
\end{equation}
is the one-loop effective potential, with  ${\cal K}(x,x')=\left[\Box_x-V''(\phi)\right]\,\delta(x-x')$.
For $\lambda=0$  (free theory) ${\cal K}(x,x')=-G_F^{-1}(x,x')$ is (minus)  the inverse Feynman propagator and satisfies the relation  $\int d^n x' {\cal K}(x,x') G_F((x',x'')=-\delta(x-x'')$, where the Green's function equation for the free field reads $(\Box_x- m^2) G_F(x,x')= -\delta(x-x')$. The integral in \eqref{eq:V1} can be conveniently worked out in momentum space:
\begin{eqnarray}\label{eq:TrV1}
Tr\,\ln{\cal K}=\Omega\,\int \frac{d^n
k}{(2\pi)^n}\,\ln\left(k^2+V''(\phi)\right)\,,
\end{eqnarray}
where e.g.  $V''(\phi)=m^2+\frac12\lambda\phi^2$ for the general Higgs potential.  Inserting  \eqref{eq:TrV1} in \eqref{eq:V1} the spacetime
volume $\Omega$  cancels  and after a simple calculation of the integral (upon Wick rotation into Euclidean momentum)  with $\lambda=0$,  one finally retrieves  Eq.\,\eqref{Vacfree2}).

How to get now a finite (though not yet necessarily physical) value of the ZPE?  Let us just follow
the renormalization program. Recall  the Einstein-Hilbert (EH) action (with CC and matter) from which Einstein's gravitational field equations
are derived:
\begin{equation}\label{eq:EHm}
S_{\rm EH+m}=  \frac{1}{16\pi G_N}\int d^4 x \sqrt{-g}\, R  -  \int d^4 x \sqrt{-g}\, \rL + S_{\rm m}\,.
\end{equation}
The matter action $S_m$ is generic at this point, but it will be specified in the next section. The (constant) term $\rL$ has dimension of energy density.  Because it is unrelated to the matter part (fully contained in $S_m$), $\rL$ is usually called the vacuum energy density. However, we will not call it that way here since it is not yet the physical vacuum energy density, $\rv$,  as we shall see.  For us the term $\rL$ is  just a bare parameter of the EH action, as the gravitational coupling  $G_N$ itself. The physical values can only be identified   after renormalizing the bare theory. The corresponding  gravitational field equations emerging from the variation of the action  \eqref{eq:EHm} can be put in the convenient form
\begin{equation} \label{FieldEq2}
\MPl^2\, G_{\mu \nu}=-\rho_\Lambda g_{\mu \nu}+T_{\mu \nu}^{\rm m}\,,
\end{equation}
where $G_{\mu\nu}=R_{\mu\nu}-(1/2) g_{\mu\nu} R$   is the Einstein tensor,  $\MPl=\mpl/\sqrt{8\pi}=1/\sqrt{8\pi G_N}$ is the reduced Planck mass,  and  $ T_{\mu \nu}^{\rm m}$  is the  stress-energy-momentum tensor, or just energy-momentum tensor (EMT) of matter. The quantity to be associated with the physically measured value of the cosmological constant, $\CC$, is  defined through $\rv=\Lambda/(8\pi G_N)$,  where $\rv$  and $G_N$ are the physical quantities. The latter, as already indicated, can only be identified after properly renormalizing the QFT calculation.  At the moment  the parameters in the action are to be understood as  bare couplings, and so none of them is physical at this stage.  Thus, $\rL$ is certainly \textit{not} the VED $\rv$.  Similarly,  if we would write  $\rL=\CC/(8\pi G_N)$  the parameter $\CC$  could \textit{not} be interpreted as the physically (measured) cosmological constant, but just as the bare cosmological term.  Now the bare parameters are  UV-divergent.  Nonetheless, to avoid cluttering we need not introduce special notations  at this point.  When necessary (as we shall see right next), the renormalized parameters will be highlighted by letting them exhibit the explicit scale dependency.

Because $\rL$  in (\ref{eq:EHm}) is  formally divergent, when we include the ZPE as a part of the full effective action we can form the sum of UV-divergent quantities $\rL+\ZPE$, which has a chance to be finite, even though not necessarily physical.  According to conventional wisdom, we may renormalize the theory
by decomposing the bare parameters, e.g.  $\rL$,  into a renormalized part plus a counterterm, $\rL=\rL(\mu)+\delta\rL$, where the splitting
depends on some arbitrary renormalization scale $\mu$. The counterterm  $\delta\rL$  depends on the regularization and
renormalization schemes, but it does not carry $\mu$-dependency. In the MS scheme, one introduces  a counterterm just killing the ``bare bone'' UV-part (the pole at
$n\rightarrow 4$). In the slightly modified $\overline{\rm MS}$
scheme\,\cite{Collins84,Sterman93} one collects in it also some additive constants. Specifically, the
counterterm for the latter is
\begin{equation}\label{deltaMSB}
{\delta}\rL^{\overline{\rm
MS}}=\frac{m^4\,\hbar}{4\,(4\pi)^2}\,\left(\frac{2}{4-n}+\ln
4\pi-\gamma_E\right)\,.
\end{equation}
Since the effective action of the bare theory must equal that of the renormalized
theory (in whatever renormalization scheme), we must have
$\rL+\ZPE=\rL(\mu)+\ZPE(\mu)$, and hence the $\mu$-dependency must cancel on the \textit{r.h.s.}
The  $\overline{\rm MS}$-renormalized one-loop result therefore reads
\begin{equation}\label{renormZPEoneloop}
V^{(1)}_{\rm ZPE}(\mu)=\hbar\,V_P^{(1)}+\delta\rL^{\overline{\rm MS}}=\frac{m^4\,\hbar}{4\,(4\,\pi)^2}\,\left(\ln\frac{m^2}{\mu^2}-\frac32\right)\,,
\end{equation}
which is perfectly finite. As previously warned, this does not necessarily mean that the obtained expression is a physical
result, but at least is a meaningful, finite,  expression.  Now as it is well-known from renormalization theory\,\cite{Collins84}, the price for
the finiteness of any renormalized result is its necessary dependence on an
arbitrary mass scale,  $\mu$ in this case.  In point of fact, this is the essence of the renormalization group\,\cite{Collins84,Sterman93,Brown92}}.  However,  as noted before, scale dependencies (including explicit and implicit)  must cancel out in the full effective action of the theory since by construction we started from the bare theory,
which is of course $\mu$-independent (RG-invariant). Hence, explicit changes of $\mu$ (within a given renormalization scheme) in one sector of the theory must be compensated by explicit and/or implicit changes of $\mu$ in another sector (e.g. from the implicit running of the couplings and fields).  Thus,  despite of the fact that both pieces, the renormalized ZPE
(\ref{renormZPEoneloop}) and the renormalized parameter $\rL(\mu)$ separately depend on the renormalization scale
$\mu$, the sum
\begin{equation}\label{renormZPEoneloop2}
\rVu=\rL+V^{(1)}_{\rm ZPE}=\rL(\mu)+V^{(1)}_{\rm ZPE}(\mu)=\rL(\mu)+\frac{m^4\,\hbar}{4\,(4\,\pi)^2}\,\left(\ln\frac{m^2}{\mu^2}-\frac32\right)
\end{equation}
does \emph{not}. The formal statement is that the logarithmic derivative
$d/d\ln\mu=\mu d/d\mu$ of the renormalized \textit{r.h.s.} of Eq.\,\eqref{renormZPEoneloop2} must be zero since the corresponding (bare) \textit{l.h.s.} obviously satisfies such condition:  $d\rV^{(1)}/d\ln\mu=0$.  It immediately follows that $\mu\frac{d\rL(\mu)}{d\mu}=\frac{\hbar\,m^4}{2\,(4\pi)^2}=\beta_{\rL}^{(1)}$, which is just the RG equation for $\rL(\mu)$.
This explains why $\beta_{\rL}^{(1)}$  in Eq,\,\eqref{Vacfree2} was called  the one-loop coefficient of the $\beta$-function for the coupling $\rL$.   However, let us no be mistaken, the previous equation
for the formal renormalized parameter $\rL(\mu)$ is by no means the RG equation for the VDE.   A lot more of work is needed yet, including a change of renormalization scheme! (cf. Sec.\,\ref{Sec:EffectiveActionl}).

The RG tells us something useful about the explicit $\mu$-dependency
affecting incomplete structures of the effective action. Indeed, while we
may not know the full structure of the effective action in a particular
complicated situation, our educated guess on associating $\mu$ with some
relevant dynamical variable of the system can furnish relevant information, similarly as we do in particle physics.  Such is e.g. the case
with the  effective charges or couplings (so-called ``running coupling constants'') in  QED
or QCD.  In any of these theories, a given renormalized gauge coupling $g=g(\mu)$  is explicitly
$\mu$-dependent even though the full effective action or $S$-matrix
element is not.  This feature can be useful to associate $\mu$ with a characteristic energy/momentum scale of the problem and hence estimate the leading quantum  effects of the process\cite{Collins84,Sterman93,Brown92}).  However, in curved spacetime the situation is more complicated. We shall retake this problem in the next section, where we shall  see that  the VED of the expanding universe depends on the expansion rate, $H$.  In the meanwhile let us briefly address the contribution from the SSB (if present) and let us do  it once more in flat spacetime for the sake of  simplicity.

In our discussion of the one-loop calculation of the VED above we have just computed the one-loop  ZPE,  which is the $V_P^{(1)}$  piece of $V_1=V_P^{(1)}+V_\text{scal}^{(1)}(\phi)$.  Now what about $V_\text{scal}^{(1)}(\phi)$?  Coming back to Eq. \eqref{eq:TrV1} and substituting it in (\ref{eq:V1}), we may trivially split the result in
the suggestive form
\begin{equation}\label{V3}
\EP^{(1)}=\hbar\,V_1=-\,\frac{i}{2}\,\hbar\,\int\frac{d^n
k}{(2\pi)^n}\,\ln\left[k^2+m^2\right]-\,\frac{i}{2}\,\hbar\,\int\frac{d^n
k}{(2\pi)^n}\,\ln\left(\frac{k^2+V''(\phi)}{k^2+m^2}\right)\,.
\end{equation}
The first term on the \textit{r.h.s} of \eqref{V3} is independent of $\phi$ and it corresponds to the previously computed  $\ZPE$ at one loop, Eq.\,\eqref{Vacfree2}, whereas
 the second term gives precisely the one-loop correction to the
$\phi$-dependent part of the effective potential, i.e.  the desired $V_\text{scal}^{(1)}(\phi)$.  Sticking to the $\overline{MS}$ scheme in  DR to
fix the counterterms, one ends up with the well-known result for the renormalized
effective potential (\ref{EPZPE}) up to one-loop\,\cite{Coleman,BooksQFT}:
\begin{eqnarray}\label{VOVR2}
\EP(\phi)=
\frac12\,m^2(\mu)\,\phi^2+\frac{1}{4!}\,\lambda(\mu)\,\phi^4
+\hbar\,\frac{\left(V''(\phi)\right)^2}{4(4\pi)^2}\left(\ln\frac{V''(\phi)}{\mu^2}-\frac32\right)\,.
\end{eqnarray}
 Since $V''(\phi=o)=m^2$, it is clear both from \eqref{V3} and \eqref{VOVR2} that we can recapture the expression of
the ZPE \,\eqref{renormZPEoneloop} from the effective potential for $\phi=0$:
\begin{eqnarray}\label{VeffRen2}
\EPR(\phi=0)=\frac{\hbar\,m^4}{4(4\pi)^2}\left(\ln\frac{m^2}{\mu^2}-\frac32\right)=V^{(1)}_{\rm ZPE}(\mu)\,.
\end{eqnarray}
This quantity must be subtracted from \eqref{VOVR2} if we are interested in  $V_\text{scal}^{(1)}(\phi)$ only.

We have remarked in \eqref{VOVR2} the implicit $\mu$-dependence of the mass and coupling. Together
with the explicit $\mu$-dependent parts of the effective potential, this
is necessary for the RG-invariance. However, the full effective potential \eqref{EPZPE} is actually
\textit{not} RG-invariant (contrary to some
inaccurate statements in the literature). It becomes so only after we
add up to it the renormalized parameter  $\rL(\mu)$:
\begin{equation}\label{FullRGequation}
\rho_{\rm tot}=\rL(m(\mu), \lambda(\mu); \mu)+\EP(\phi(\mu); m(\mu), \lambda(\mu); \mu) \ \ \ \ \Rightarrow \ \ \ \ \ \mu\frac{d\rho_{\rm tot}}{d\mu}=0\,.
\end{equation}
This expression  generalizes of course Eq.\,\eqref{renormZPEoneloop2} for the case when we incorporate the $\phi$-dependent part of  the effective potential.
The RG-invariance of \eqref{renormZPEoneloop2}  is  still maintained because this part does not feel the scaling of the fields (as there are no external legs in the ZPE).  It follows that in the general case  we can split \eqref{FullRGequation} into two blocks which are independently RG-invariant:
\begin{equation}\label{RGequation1}
\mu\frac{d}{d\mu}\left[\rL(m(\mu), \lambda(\mu); \mu))+\ZPE(m(\mu),
\lambda(\mu); \mu)\right]=0
\end{equation}
and
\begin{equation}\label{RGequation2}
\mu\frac{d}{d\mu}\, V_{\rm scal}(\phi(\mu); m(\mu), \lambda(\mu);
\mu)=0\,.
\end{equation}
One may  be tempted to conceive that the VEV of  \eqref{FullRGequation},  $\langle\rho_{\rm tot}\rangle$ ,  should represent the $\overline{\rm MS}$-renormalized VED of the scalar field. In a sense it is so, if such quantity is to acquire a physical meaning in the context of QFT in flat spacetime.  However, even in that case this does not entail any relationship whatsoever of   $\langle\rho_{\rm tot}\rangle$ with cosmology and the CCP, for the vacuum energy can only have full (absolute) meaning in the gravitational context. The reason is simple: the geometry of the universe depends on its entire energy budget. Thus, even if the above relations are formally  correct and well-known by many (as I presume),  one cannot deal with the VED in cosmology along these lines since they are based on pure QFT in Minkowskian spacetime.  To start with, we need curvature  and hence we expect a dependency of the result on the expansion rate $H$ and its derivatives.  The scale $\mu$ in these expressions is purely formal and cannot be linked to any physical quantity in cosmology.  The VED in cosmology is neither the formal parameter $\rL(\mu)$ nor the VEV of the effective potential $\EP(\mu)$, not even  the RG-invariant sum of both.   Moreover, we expect that measurements of the  VED should be free of $\sim m^4$ contributions which cause the abhorrent fine tuning problem inherent to the CCP.

From the fundamental perspective of  QFT in curved spacetime, the cosmological VED is to be naturally ascribed to a vacuum renormalization effect. It is, therefore, a scale-dependent quantity, thereby  evolving with the renormalization scale. The overall scale dependencies cancel out in the full effective action, of course, which has many other scale-dependent terms, for example  the masses, couplings and fields of the classical Lagrangian part of the full quantum action.  Quite obviously, we have to go beyond the above  flat spacetime considerations, which have no implications at all on the VED and the CC in cosmology. It is surprising that sometimes these naive  flat spacetime formulas are used as an attempt to evaluate the CC, see\,\cite{JMartin2012} and references therein.  To describe the  CC and the VED in cosmology we need, first of all,  a nontrivial framework where the CC can be properly defined: hence we need Einstein's equations in curved spacetime. The full  effective action, on the other hand, comprises the Einstein-Hilbert term, but it must involve additional geometric structures as well (e.g. the higher derivatives terms). The VED  can only be meaningfully if defined in this context.  In it,  the parameter $\rL$ is only one more piece;  but, as we have seen,  it is indispensable  to renormalize the ZPE,  even in Minkowski space.  This will also be so  in curved spacetime, but things will get a bit  more complicated.  In the presence of SSB, additional complications add up to the calculation of the total VED.   Heretofore we have discussed both  the ZPE and the SSB  in the simpler flat spacetime context. But this simplification precludes us  from establishing a physical connection between VED and CC.   In what follows we jump to QFT in curved spacetime and  reexamine the situation.  However, we will  limit ourselves  to discuss the  renormalization of the ZPE, or to be more precise of the combined quantity $\rL+{\rm ZPE}$,  which in itself is already quite involved in curved space. Even though one can still use  the MS scheme,  we shall abandon it as a renormalization prescription because it definitely does not produce physical results in this context  and leads to the awful fine-tuning difficulties around the CCP owing to  the left over  $\sim m^4$ terms in the result\,\cite{KohriMatsui2017}.  Unavoidably, we are led to use a more physical renormalization scheme.  From now on we set $\hbar=1$, except in a particular instance.

\section{Non-minimally coupled scalar field in FLRW background}\label{sec:EMT}

We take up the discussion of the ZPE for a scalar field, now  within  QFT in curved spacetime. Calculations here are harder, of course.  In addition, in view of the limitations of the MS scheme to provide a  more physical renormalization of the VED, we shall adopt an alternative scheme (a variant of the so-called adiabatic regularization and renormalization method\,\cite{BirrellDavies82,ParkerToms09}).  Furthermore, to make our task not exceedingly burdensome, we shall hereafter assume that there is no SSB, hence no scalar field self interactions ($\lambda=0$). In this case,   $m^2>0$ is the physical mass squared of the field. The ZPE part in a curved background is already sufficiently demanding even when the scalar field does not couple to itself.  This means we shall focus on just the one-loop piece of $\ZPE$  in Eq.\,\eqref{EPZPE}. Within this, more restricted,  setup we will tackle  the  calculation of the energy density of the vacuum fluctuations of a quantized scalar field  in FLRW spacetime.  For the sake of completeness, we admit the presence of a  non-minimal coupling $\xi$ of $\phi$ to gravity, despite  it is not mandatory for renormalizability in this case.  Since there are no self couplings in our calculation, there are no ZPE loops beyond one, and hence the one-loop contribution that we are addressing here, $\ZPE=\hbar\,V_P^{(1)}$, is in fact the full ZPE result.  The calculation entails renormalization subtleties since we meet, of course,  UV-divergent integrals. As noted, conventional approaches  such as e.g. the MS scheme,  lead to  the very same  conflictive terms $\sim m^4$  as those previously met in Minkowski space\,\cite{JSPRev2013}.   We shall therefore eschew unsuccessful renormalization methods of this sort in the cosmological arena. We shall use instead the adiabatic renormalization procedure (ARP)\cite{BirrellDavies82,ParkerToms09,Fulling89} to renormalize the VED,  although implemented off-shell via the  approach  defined  in \cite{CristianJoan2020,CristianJoan2022} and introduced for running couplings in \cite{FerreiroNavarroSalas}.

\subsection{Energy-momentum tensor}\label{Sec:FieldEquations}

The curved space  EMT of matter follows  from the metric functional variation of the matter action:
\begin{equation}\label{eq:deltaTmunu2}
  T_{\mu \nu}^{\rm m}=-\frac{2}{\sqrt{-g}}\frac{\delta S_{\rm m}}{\delta g^{\mu\nu}}\,.
\end{equation}
At this point we may assume for simplicity that there is only one (matter) field contribution to the EMT  on the right hand side of \eqref{FieldEq2} in the form of  a real scalar field, $\phi$. Such contribution will be denoted $T_{\mu \nu}^{\phi}$.   We need not include  the incoherent matter contributions  from dust and radiation at this point.  They can be added a posteriori  without altering the pure  QFT discussion on  which we shall focus next.  The part of the action involving $\phi$ will be taken as  follows:
\begin{equation}\label{eq:Sphi}
  S[\phi]=-\int d^4x \sqrt{-g}\left(\frac{1}{2}g^{\mu \nu}\partial_{\mu} \phi \partial_{\nu} \phi+\frac{1}{2}(m^2+\xi R)\phi^2 \right)\,.
\end{equation}
The non-minimal coupling of $\phi$ to gravity is assumed arbitrary.  It is well-known that for  $\xi=1/6$, the massless ($m=0$)  action is then (locally) conformal invariant in $n=4$ spacetime dimensions .   As noted, we also  assume  that $\phi$   has no effective potential.  In this study, we wish to concentrate on the  ZPE  of $\phi$ only.

The classical  EMT ensues from the metric functional derivative of  the action \eqref{eq:Sphi}:
\begin{equation}\label{EMTScalarField}
\begin{split}
T_{\mu \nu}^{\phi}=&-\frac{2}{\sqrt{-g}}\frac{\delta S[\phi]}{\delta g^{\mu\nu}}= (1-2\xi) \partial_\mu \phi \partial_\nu\phi+\left(2\xi-\frac{1}{2} \right)g_{\mu \nu}\partial^\sigma \phi \partial_\sigma\phi\\
& -2\xi \phi \nabla_\mu \nabla_\nu \phi+2\xi g_{\mu \nu }\phi \Box \phi +\xi G_{\mu \nu}\phi^2-\frac{1}{2}m^2 g_{\mu \nu} \phi^2.
\end{split}
\end{equation}
 Varying the action \eqref{eq:Sphi} with respect to $\phi$ we find the  Klein-Gordon (KG) equation in curved spacetime:
\be\label{eq:KG}
(\Box-m^2-\xi R)\phi=0\,,
\ee
where $\Box\phi=g^{\mu\nu}\nabla_\mu\nabla_\nu\phi=(-g)^{-1/2}\partial_\mu\left(\sqrt{-g}\, g^{\mu\nu}\partial_\nu\phi\right)$. The FLRW line element for spatially flat three-dimensional geometry can be written in conformal coordinates as  $ds^2=a^2(\tau)\eta_{\mu\nu}dx^\mu dx^\nu$, where  $\eta_{\mu\nu}={\rm diag} (-1, +1, +1, +1)$ is the Minkowski  metric in our conventions.  Differentiation with respect to the conformal time, $\tau$, will be denoted  with a prime $^\prime\equiv d/d\tau$,  whereas dot  indicates derivative with respect to the cosmic time ( $\dot{}\equiv d/dt$). The  Hubble rate in conformal time  $\mathcal{H}(\tau)\equiv a^\prime /a$ is related to the Hubble rate in cosmic time through $\mathcal{H}(\tau)=a  H(t)$,  in which  $H(t)=\dot{a}/a$.

 The KG equation \eqref{eq:KG} in conformally flat coordinates takes the form
\begin{equation}\label{eq:KGexplicit}
 \phi''+2\cH\phi'-\nabla^2\phi+a^2(m^2+\xi R)\phi=0\,,
\end{equation}
where we used the curvature scalar  of FLRW spacetime:  $R=6a^{\prime\prime}/a^3$ .
The separation of variables  in these coordinates, namely $\phi(\tau,x)\sim \int d^3k \ A_{\bf k}\psi_k({\bf x})\phi_k(\tau)+cc$, can  be achieved with $\psi_k(x)=e^{i{\bf k\cdot x}}$. However, in contrast to the Minkowski case we cannot take  $\phi_k(\tau)=e^{\pm i\omega_k \tau}$ since the mode frequencies are not constant anymore.  The form of the  modes $\phi_k(\tau)$  in the curved spacetime case are determined by the KG equation.  Starting from the Fourier expansion with separated space and time variables
\begin{equation}\label{FourierModes}
\phi(\tau,{\bf x})=\int\frac{d^3{k}}{(2\pi)^{3/2}} \left[ A_\bk e^{i{\bf k\cdot x}} \phi_k(\tau)+A_\bk^\ast e^{-i{\bf k\cdot x}} \phi_k^*(\tau) \right]
\end{equation}
(in which  $A_\bk $  and their complex conjugates $A_\bk^\ast$  are the classical Fourier coefficients) and substituting it into \eqref{eq:KGexplicit} the functions $\phi_k(\tau)$ are determined by solving the nontrivial differential equation
\begin{equation}\label{eq:KGFourier}
 \phi_k''+2\cH\phi'_k+\left(\omega_k^2(m)+a^2\xi R\right)\phi_k=0\,,
\end{equation}
where $\omega_k^2(m)\equiv k^2+a^2 m^2$. The mode functions depend only on the modulus $k\equiv|\bk|$ of the comoving momenta  $(\tilde{k}=k/a$ being the physical ones). The frequencies are in general functions of the time-evolving scale factor $a=a(\tau)$, and this makes the particle interpretation harder. But we do not aim at any particle interpretation here, we target at an interpretation in terms of field observables and ultimately in terms of the EMT constructed from QFT in curved spacetime (see next sections).  If we perform the change of field mode variable  $\phi_k=\varphi_k/a$   the above equation simplifies to
$
\varphi_k^{\prime \prime}+\left(\omega_k^2(m)+a^2\,(\xi-1/6)R)\right)\varphi_k=0\,.
$
For conformally invariant matter ($m=0$ and $\xi=1/6$),  such equation boils down to  a simple form which admits  positive- and negative-energy solutions  $e^{- ik\tau}$  and $e^{+ ik\tau}$. In the massless case with minimal coupling ($\xi=0$) analytical solutions also exist. But  for $m\neq0$ and/or $\xi\neq 1/6$ no analytic solution of \eqref{eq:KGFourier} is available, and this leads us to perform  a WKB (Wentzel-Kramers-Brillouin)  expansion of the solution.  Before tackling that method, let us first  briefly review  the quantization of a non-minimally coupled scalar field $\phi$.

 \subsection{Quantum fluctuations}\label{sec:AdiabaticVacuum}

To construct  $\langle T_{\mu \nu}^\phi \rangle\equiv \langle  0 |T_{\mu \nu}^\phi |0 \rangle$, i.e.  the VEV  of the  EMT from quantum matter fields in a curved background, we need to account for the quantum fluctuations of the fields themselves. In our simplified treatment, we have just  the scalar field $\phi$  and therefore we consider the expansion of the field around its background value $\phi_b$:
\begin{equation}
\phi(\tau,x)=\phi_b(\tau)+\delta\phi (\tau,x). \label{ExpansionField}
\end{equation}
One starts defining  a suitable vacuum state,  $ |0 \rangle$, called adiabatic vacuum\cite{Bunch1980}. The  VEV of $\phi$  is  identified with the background value,  $\langle 0 | \phi (\tau, x) | 0\rangle=\phi_b (\tau)$, whereas the VEV of the fluctuation is zero:  $\langle  \delta\phi  \rangle\equiv \langle 0 | \delta\phi | 0\rangle =0$. This is not the case for the VEV of the bilinear products of fluctuations, e.g. $\langle \delta\phi^2 \rangle\neq0$.  It is convenient to decompose  $\langle T_{\mu \nu}^\phi \rangle=\langle T_{\mu \nu}^{\phi_b} \rangle+\langle T_{\mu \nu}^{\delta_\phi}\rangle$, where
$\langle T_{\mu \nu}^{\phi_{b}} \rangle =T_{\mu \nu}^{\phi_{b}} $
is the  contribution  from the classical background part,  whereas the important part $\langle T_{\mu \nu}^{\delta\phi}\rangle$  is  the genuine vacuum contribution from the field fluctuations $\delta\phi$.  In curved spacetime,  the term  $\rho_\Lambda$   is a full-fledged  part of the vacuum action  \eqref{eq:EHm} and with obvious connections with the cosmological term after renormalization.  The complete vacuum EMT  is indeed  the sum
\begin{equation}\label{EMTvacuum}
\langle T_{\mu \nu}^{\rm vac} \rangle=-\rho_\Lambda g_{\mu \nu}+\langle T_{\mu \nu}^{\delta \phi}\rangle\,.
\end{equation}
Let us treat  the vacuum as a perfect fluid, namely with an EMT of the form
\begin{equation}\label{eq:VaccumIdealFluid}
\langle T_{\mu\nu}^{{\rm vac}}\rangle=\Pv g_{\mu \nu}+\left(\rv+\Pv\right)u_\mu u_\nu\,,
\end{equation}
where $u^\mu$ is the $4$-velocity of the observer. No assumption is made, however,  on the equation of state of the vacuum.  In conformal coordinates in  the comoving cosmological (FLRW)  frame,  we have $u^\mu=(-a,0,0,0)$ and hence  $u_\mu=(a,0,0,0)$. Taking the $00th$-component of  \eqref{eq:VaccumIdealFluid},  the relation  $\langle T_{00}^{{\rm vac}}\rangle=-a^2 \Pv+\left(\rv+\Pv\right) a^2=a^2\rho_{\rm vac}$ follows, irrespective of $\Pv$.  Whence
\begin{equation}\label{unRenVDE}
\rv= \frac{\langle T_{00}^{\rm vac}\rangle}{a^2}= \rho_\Lambda+\frac{\langle T_{00}^{\delta \phi}\rangle}{a^2}\,,
\end{equation}
where  we have inserted  the $00th$-component of \eqref{EMTvacuum} in the second equality.
We can see that the total vacuum part  receives contributions from both the cosmological term in the action as well as from the quantum fluctuations  of the field (i.e. the ZPE).  This is nothing but the curved space counterpart (in conformally flat coordinates) to the Minkowskian spacetime relation \eqref{renormZPEoneloop2}. However,  in the curved space case the above  relation \eqref{unRenVDE} is still at the stage  of bare quantities and hence they are formally UV-divergent. The theory must be renormalized.   For this it will be necessary to use an appropriate subtraction prescription, as we shall see later on.

At the moment, to proceed we have to solve for  the field modes in the curved background. Let us redefine the field  $\phi=\varphi/a$, where $a$ is the scale factor. Denoting the frequency modes of the fluctuating part  $\delta\varphi$ by   $h_k(\tau)$, we can write
\begin{equation}\label{FourierModesFluc}
\delta \varphi(\tau,{\bf x})=\int \frac{d^3{k}}{(2\pi)^{3/2}} \left[ A_\bk e^{i{\bf k\cdot x}} h_k(\tau)+A_\bk^\dagger e^{-i{\bf k\cdot x}} h_k^*(\tau) \right]\,,
\end{equation}
where  $A_\bk$ and  $A_\bk^\dagger $ are now  the (time-independent) annihilation and creation operators. They satisfy the commutation relations
\begin{equation}\label{CommutationRelation}
[A_\bk, A_\bk'^\dagger]=\delta({\bf k}-{\bf k'}), \qquad [A_\bk,A_ \bk']=0.
\end{equation}
The frequency modes of the fluctuations, $h_k(\tau)$, satisfy the  differential equation
\begin{equation}\label{eq:ODEmodefunctions}
h_k^{\prime \prime}+\Omega_k^2(\tau) h_k=0\ \ \ \ \ \ \ \ \ \ \ \Omega_k^2(\tau) \equiv\omega_k^2(m)+a^2\, (\xi-1/6)R\,.
\end{equation}
The solution of that equation in the general case requires a recursive self-consistent iteration,  the  WKB expansion.  One starts from
\begin{equation}\label{eq:phaseIntegral}
h_k(\tau)=\frac{1}{\sqrt{2W_k(\tau)}}\exp\left(-i\int^\tau W_k(\tilde{\tau})d\tilde{\tau} \right)\,,
\end{equation}
where the  normalization $1/\sqrt{2W_k(\tau)}$ insures the Wronskian condition
$ h_k^{} h_k^{*\prime}- h_k^*h_k^\prime=i$,
which warrants the standard equal-time commutation relations for the fields.
Functions $W_k$ in the above ansatz obey the (non-linear) equation
\begin{equation} \label{WKBIteration}
W_k^2(\tau)=\Omega_k^2(\tau) -\frac{1}{2}\frac{W_k^{\prime \prime}}{W_k}+\frac{3}{4}\left( \frac{W_k^\prime}{W_k}\right)^2\,,
\end{equation}
which is amenable to be  solved using the WKB expansion. The latter is applicable only for large $k$, therefore  short wave lengths (as e.g. in geometrical Optics),  and weak gravitational fields.  The mode functions $h_k(\tau)$ are no longer of the form  $\varphi_k(\tau)=e^{\pm i\omega_k\tau}$,  hence particles with definite frequencies cannot be strictly defined in a curved background.  An approximate Fock space interpretation is still feasible if the  vacuum is defined as the quantum state which is annihilated by all the operators $A_{\bf k}$ of the above Fourier expansion. Such an  adiabatic vacuum is crucial to  attain a physical interpretation of the results in terms of a properly renormalized EMT.

\subsection{WKB expansion of the mode functions}\label{sec:WKB}

 The  WKB expansion is very helpful in this context\,\cite{BirrellDavies82,ParkerToms09}.  It eases the implementation of  the ARP  in terms of  subtracted integrals which become UV-finite and hence it prompts a direct renormalization of the theory.  In the work\,\cite{CristianJoan2020}, the WKB expansion of the field modes was performed  up to $4th$;   subsequently it was extended up to $6th$ adiabatic order  in \cite{CristianJoan2022}.  The amount of calculational effort of the second step is rather formidable,  but it proves necessary to  study certain aspects of the on-shell versus  off-shell renormalized theory. Here we will limit ourselves to describe the results up to $4th$ order, which is enough to renormalize the theory off-shell (see the aforementioned papers for more details).   The counting of adiabatic orders in the WKB expansion follows the number of time derivatives. Thus:  $k^2$ and $a$ are of adiabatic order $0$;  $a^\prime$ and $\mathcal{H}$  of adiabatic order 1;  $a^{\prime \prime},a^{\prime 2},\mathcal{H}^\prime$ and $\mathcal{H}^2$ as well as $R$ ,are of adiabatic order $2$. Each additional derivative increases the adiabatic order  by one unit.   The expansion collects the different adiabatic orders:
\begin{equation}\label{WKB}
W_k=\omega_k^{(0)}+\omega_k^{(2)}+\omega_k^{(4)}+\omega_k^{(6)}\cdots
\end{equation}
General covariance precludes the odd adiabatic orders.
Following \cite{CristianJoan2020,CristianJoan2022}, we consider an off-shell procedure to renormalize the VED  in which the frequency $\omega_k$ of a given mode  is defined not at the mass $m$ of the particle but at an arbitrary mass scale $M$:
$\omega_k\equiv\omega_k(\tau, M)\equiv \sqrt{k^2+a^2(\tau) M^2}$.   It is akin to the method of  \cite{FerreiroNavarroSalas} to renormalize the couplings, but with some differences explained in \cite{CristianJoan2022} and of course applied to the VED in our case.  We avoid using the notation  $\mu$ here to avoid connotations with 't Hooft's  mass unit in the MS scheme. Also because $\mu$ will be used in combination with $M$ in Sec.\,\ref{Sec:EffectiveActionl}.
The second and  fourth order adiabatic terms can be computed with the following result:
\begin{equation}
\begin{split}
\omega_k^{(0)}&= \omega_k\,,\\
\omega_k^{(2)}&= \frac{a^2 \Delta^2}{2\omega_k}+\frac{a^2 R}{2\omega_k}(\xi-1/6)-\frac{\omega_k^{\prime \prime}}{4\omega_k^2}+\frac{3\omega_k^{\prime 2}}{8\omega_k^3}\,,\\
\omega_k^{(4)}&=-\frac{1}{2\omega_k}\left(\omega_k^{(2)}\right)^2+\frac{\omega_k^{(2)}\omega_k^{\prime \prime}}{4\omega_k^3}-\frac{\omega_k^{(2)\prime\prime}}{4\omega_k^2}-\frac{3\omega_k^{(2)}\omega_k^{\prime 2}}{4\omega_k^4}+\frac{3\omega_k^\prime \omega_k^{(2)\prime}}{4\omega_k^3}\,.
\end{split}\label{WKBexpansions1}
\end{equation}
The  term  $\Delta^2\equiv m^2-M^2$  must be categorized as being of adiabatic order 2 since it appears consistently so in the WKB expansion\footnote{In Sec.\,\ref{Sec:EffectiveActionl} we shall further justify this procedure in the effective action formalism.}.  For $M=m$  (viz.  $\Delta = 0$) the on-shell expansion is recovered and corresponds to the usual ARP procedure\cite{BirrellDavies82,ParkerToms09}.   As noted, we will not report here on the cumbersome  $6th$-order  formulas\,\cite{CristianJoan2022}.   A crucial point must be  emphasized:  the adiabatic expansion ultimately becomes an expansion in powers of $\mathcal{H}$ and its time derivatives since
\begin{equation}\label{omegak0}
\omega_k^\prime=a^2\mathcal{H}\frac{M^2}{\omega_k}, \qquad\omega_k^{\prime \prime}=2a^2\mathcal{H}^2\frac{M^2}{\omega_k}+a^2\mathcal{H}^\prime \frac{M^2}{\omega_k}-a^4\mathcal{H}^2\frac{M^4}{\omega_k^3}\,.
\end{equation}

\section{Computing the ZPE in curved spacetime }\label{eq:RegZPE}

Equipped with the necessary tools and concepts to efficiently compute and correctly interpret the zero-point energy brought about  by the quantum vacuum fluctuations in the FLRW background, we may now insert  the split form of the quantum field $\phi$  as given in (\ref{ExpansionField}) in the  expression of the EMT,  Eq.\,\eqref{EMTScalarField}, and select the fluctuating parts  $\delta\phi$. Recall, however, that only the quadratic fluctuations in  $\delta\phi$ survive since $ \langle 0 | \delta\phi | 0\rangle =0$. The  ZPE  is just the VEV of the  $00$-component of the EMT:
\begin{equation}\label{EMTInTermsOfDeltaPhi}
\begin{split}
\langle T_{00}^{\delta \phi}\rangle \equiv&  \langle  0 |T_{00}^{\delta\phi} |0 \rangle=\left\langle \frac{1}{2}\left(\delta\phi^{\prime}\right)^2+\left(\frac{1}{2}-2\xi\right)\left(\nabla\delta \phi\right)^2+6\xi\mathcal{H}\delta \phi \delta \phi^\prime\right.\\
&\left.-2\xi\delta\phi\,\nabla^2\delta\phi+3\xi\mathcal{H}^2\delta\phi^2+\frac{a^2m^2}{2}(\delta\phi)^2 \right\rangle\,,
\end{split}
\end{equation}
where $\delta\phi'$ is the fluctuation of the derivative of $\phi$  with respect to conformal time.   Next we substitute the Fourier expansion of $\delta\phi=\delta\varphi/a$, as given in \eqref{FourierModesFluc},  into Eq.\,\eqref{EMTInTermsOfDeltaPhi} and use the commutation relations \eqref{CommutationRelation}. Upon symmetrizing  the operator field products $\delta\phi \delta\phi^\prime$ with respect to the creation and annihilation operators and jumping to Fourier space by integrating  $\int\frac{d^3k}{(2\pi)^3}(...) $  actual calculation leads to\cite{CristianJoan2020,CristianJoan2022}:
\begin{equation}\label{EMTFluctuations}
\begin{split}
\langle T_{00}^{\delta \phi (0-4)} \rangle & =\frac{1}{8\pi^2 a^2}\int dk k^2 \left[ 2\omega_k+\frac{a^4M^4 \mathcal{H}^2}{4\omega_k^5}-\frac{a^4 M^4}{16 \omega_k^7}(2\mathcal{H}^{\prime\prime}\mathcal{H}-\mathcal{H}^{\prime 2}+8 \mathcal{H}^\prime \mathcal{H}^2+4\mathcal{H}^4)\right.\\
&+\frac{7a^6 M^6}{8 \omega_k^9}(\mathcal{H}^\prime \mathcal{H}^2+2\mathcal{H}^4) -\frac{105 a^8 M^8 \mathcal{H}^4}{64 \omega_k^{11}}\\
&+\left(\xi-\frac{1}{6}\right)\left(-\frac{6\mathcal{H}^2}{\omega_k}-\frac{6 a^2 M^2\mathcal{H}^2}{\omega_k^3}+\frac{a^2 M^2}{2\omega_k^5}(6\mathcal{H}^{\prime \prime}\mathcal{H}-3\mathcal{H}^{\prime 2}+12\mathcal{H}^\prime \mathcal{H}^2)\right. \\
& \left. -\frac{a^4 M^4}{8\omega_k^7}(120 \mathcal{H}^\prime \mathcal{H}^2 +210 \mathcal{H}^4)+\frac{105a^6 M^6 \mathcal{H}^4}{4\omega_k^9}\right)\\
&+\left. \left(\xi-\frac{1}{6}\right)^2\left(-\frac{1}{4\omega_k^3}(72\mathcal{H}^{\prime\prime}\mathcal{H}-36\mathcal{H}^{\prime 2}-108\mathcal{H}^4)+\frac{54a^2M^2}{\omega_k^5}(\mathcal{H}^\prime \mathcal{H}^2+\mathcal{H}^4) \right)
\right]\\
&+\frac{1}{8\pi^2 a^2} \int dk k^2 \left[  \frac{a^2\Delta^2}{\omega_k} -\frac{a^4 \Delta^4}{4\omega_k^3}+\frac{a^4 \mathcal{H}^2 M^2 \Delta^2}{2\omega_k^5}-\frac{5}{8}\frac{a^6\mathcal{H}^2 M^4\Delta^2}{\omega_k^7} \right.\\
& \left. +\left( \xi-\frac{1}{6} \right) \left(-\frac{3a^2\Delta^2 \mathcal{H}^2}{\omega_k^3}+\frac{9a^4 M^2 \Delta^2 \mathcal{H}^2}{\omega_k^5}\right)\right]+\dots,
\end{split}
\end{equation}
Only even powers of $\cal H$ remain in the final result, as we expected.
It is apparent that the above expression for the ZPE is UV-divergent.  The  Minkowskian piece contained in it is also divergent.  Indeed,  taking  $a=1$ ($\mathcal{H}=0)$ in  the on-shell limit ($M=m$, hence $\Delta^2=0$)  we find the standard result obtained previously in \eqref{eq:ZPE1loop}:
\begin{equation}\label{eq:Minkoski}
  \left.\langle T_{00}^{\delta \phi}\rangle\right|_{\rm Minkowski}=\frac{1}{4\pi^2}\int dk k^2 \left(\hbar\,\omega_k\right) =  \int\frac{d^3k}{(2\pi)^3}\,\left(\frac12\,\hbar\,\omega_k\right)\,,
\end{equation}
where $\hbar$ has been  restored here just to ease the interpretation.   The result \eqref{eq:Minkoski} is  (quartically) UV-divergent.   As reviewed in Sec.\,\ref{Sec:FlatSpaceAnalysis}, usual attempts at regularizing/renormalizing this expression  (e.g. through the MS scheme  by  subtracting the pole against the bare $\rL$ term  in the action \eqref{eq:EHm})  end up with the unfathomable  fine-tuning  problem,  perhaps the most excruciating  aspect of the CCP -- see e.g.\cite{JSPRev2013} and references therein.  We will certainly take  a very different road at this crucial  juncture.

\section{VED  in FLRW spacetime: the running vacuum model (RVM)}\label{sec:RenormEMT}

The VED in the expanding FLRW universe can be viewed as emerging from  a cosmic Casimir device in which the parallel plates  move apart (“expand”)\cite{JSPRev2013}.
Although the  total vacuum energy density is not a physical quantity (it is not even definable in a consistent way), the distinctive  effect produced by the presence of the plates and their increasing separation with time, is accessible to our local measurements, as there is a distinctive local curvature $R$ as compared to Minkowskian spacetime that  is changing with the expansion.  The measurable VED must be the result of purely geometric contributions proportional to $R$, $R^2$, $R^{\mu\nu}R_{\mu\nu}$ etc., hence to $H^2$ and $\dot{H}$ (including higher powers of these quantities in the early Universe).

Following\cite{CristianJoan2020, CristianJoan2022}, we adhere to  a subtraction prescription for the VEV of the EMT that is carried out at an arbitrary mass scale $M$, which plays the role of renormalization point.
Taking into account that the only adiabatic orders that are divergent in the case of the EMT are the  first four ones (in $n=4$ spacetime dimensions), the subtraction at the scale $M$ is performed only up to the fourth adiabatic order. The on-shell value of the EMT can be computed of course at any order. The terms beyond the 4th order are  finite.
Therefore, the renormalization prescription that we adopt for the  EMT takes on the form
\begin{eqnarray}\label{EMTRenormalized}
\langle T_{\mu\nu}^{\delta \phi}\rangle_{\rm Ren}(M)&=&\langle T_{\mu\nu}^{\delta \phi}\rangle(m)-\langle T_{\mu\nu}^{\delta \phi}\rangle^{(0-4)}(M)\,.
\end{eqnarray}
Let us apply this procedure to the ZPE part of the EMT, as given by  Eq.\,\eqref{EMTFluctuations}. We shall now distinguish explicitly  between the off-shell energy mode $\omega_k(M)=\sqrt{k^2+a^2 M^2}$  (formerly denoted just as $\omega_k$) and the on-shell one  $\omega_k(m)=\sqrt{k^2+a^2 m^2}$.  Lengthy but  straightforward calculations from equations  \eqref{EMTFluctuations} and \eqref{EMTRenormalized}  lead to the compact result up to $4th$ adiabatic order:
\begin{equation}\label{Renormalized2}
\begin{split}
&\langle T_{00}^{\delta \phi}\rangle^{(0-4)}_{\rm Ren}(M)
=\frac{a^2}{128\pi^2 }\left(-M^4+4m^2M^2-3m^4+2m^4 \ln \frac{m^2}{M^2}\right)\\
&-\left(\xi-\frac{1}{6}\right)\frac{3 \mathcal{H}^2 }{16 \pi^2 }\left(m^2-M^2-m^2\ln \frac{m^2}{M^2} \right)+\left(\xi-\frac{1}{6}\right)^2 \frac{9\left(2  \mathcal{H}^{\prime \prime} \mathcal{H}- \mathcal{H}^{\prime 2}- 3  \mathcal{H}^{4}\right)}{16\pi^2 a^2}\ln \frac{m^2}{M^2}\,.
\end{split}
\end{equation}
As mentioned, the on-shell value can be computed at any order (with the limitations of an asymptotic expansion) and all these additional terms are finite.  For instance, the $6th$-order terms have been computed explicitly  in \cite{CristianJoan2022}.
 The above formula for the renormalized vacuum fluctuations is not yet the VED, however.  As prescribed  in \eqref{EMTvacuum},  we must  include as well  the contribution from the parameter $\rL$ in the Einstein-Hilbert action \eqref{eq:EHm}. Phrased in terms of renormalized parameters,
\begin{equation}\label{RenEMTvacuum}
\langle T_{\mu\nu}^{\rm vac}\rangle_{\rm Ren}(M)=-\rho_\Lambda (M) g_{\mu \nu}+\langle T_{\mu \nu}^{\delta \phi}\rangle_{\rm Ren}(M)\,.
\end{equation}
The sought-for  renormalized VED  is just the renormalized analogue of \eqref{unRenVDE} ensuing from \eqref{RenEMTvacuum} and the condition of perfect fluid. It reads
\begin{equation}\label{RenVDE}
\rho_{\rm vac}(M)= \frac{\langle T_{00}^{\rm vac}\rangle_{\rm Ren}(M)}{a^2}=\rho_\Lambda (M)+\frac{\langle T_{00}^{\delta \phi}\rangle_{\rm Ren}(M)}{a^2}\,.
\end{equation}
  As we shall see in a moment, the above expression has the necessary properties to cure the fine tuning disease.  First of all, we use the previous formulas to write it out in detail:
\begin{equation}\label{RenVDEexplicit}
\begin{split}
\rv(M)&= \rho_\Lambda (M)+\frac{1}{128\pi^2 }\left(-M^4+4m^2M^2-3m^4+2m^4 \ln \frac{m^2}{M^2}\right)\\
&-\left(\xi-\frac{1}{6}\right)\frac{3 \mathcal{H}^2 }{16 \pi^2 a^2}\left(m^2-M^2-m^2\ln \frac{m^2}{M^2} \right)\\
&+\left(\xi-\frac{1}{6}\right)^2 \frac{9\left(2  \mathcal{H}^{\prime \prime} \mathcal{H}- \mathcal{H}^{\prime 2}- 3  \mathcal{H}^{4}\right)}{16\pi^2 a^4}\ln \frac{m^2}{M^2}\,.
\end{split}
\end{equation}
The momentous  feature about this expression for the renormalized VED  is that the first two terms  (i.e. those that do  \textit{not} depend on $\cH$) exactly cancel against each other when we compute the difference between the  values of $\rv$  at two arbitrary scales, say $M$ and $M_0$:
\begin{equation}\label{eq:VEDscalesMandM0Final}
\begin{split}
&\rv(M)-\rv(M_0)=\left(\xi-\frac16\right)\frac{3\cH^2}{16\pi^2 a^2}\,\left(M^2 - M_0^{2} -m^2\ln \frac{M^{2}}{M_0^2}\right)\\
&\phantom{XXXXXXXXx}+\left(\xi-\frac16\right)^2\frac{9}{16 \pi^2 a^4}\left(\mathcal{H}^{\prime 2}-2\mathcal{H}^{\prime \prime}\mathcal{H}+3 \mathcal{H}^4 \right)\ln \frac{M^2}{M_0^{2}}\,.\\
\end{split}
\end{equation}
To verify  the precise cancellation of the mentioned terms, we start rewriting Einstein's equations\,\eqref{FieldEq2} using the renormalized parameters but  in   generalized form, which means that we  must  include the standard higher derivative (HD) tensor  $^{(1)}{\rm H}_{\mu \nu}$ (essentially the metric functional variation of the curvature squared  $R^2$, the latter being  part of the usual HD action\,\cite{CristianJoan2022}). This is,  of course,  indispensable for renormalization purposes as this term is generated by the quantum effects.  In FLRW spacetime no other higher order tensors are necessary\cite{BirrellDavies82}. Thus, using Eq.\,\eqref{RenEMTvacuum}, we have
\begin{equation}\label{eq:EqsVac2}
\MPl^2 (M) G_{\mu \nu}+\alpha(M) ^{(1)}{\rm H}_{\mu \nu}= \langle T_{\mu\nu}^{\rm vac}\rangle_{\rm Ren}(M)\,.
\end{equation}
We have written  only the vacuum part of the EMT since in doing the mentioned subtraction the background contribution of the field $\phi$  (and any other contribution) will cancel, except the ($M$-dependent) change of the vacuum EMT at the two scales. We may now subtract side by side  Eq.\,\eqref{eq:EqsVac2} at the scale values $M$ and $M_0$ and project the $00$th component of the result. For  $\langle T_{00}^{\rm vac}\rangle_{\rm Ren}(M)$ we can  use \eqref{Renormalized2}, whereas for  $ G_{00}$ and $^{(1)}{\rm H}_{00}$ in the FLRW metric  we can use the  appendices of \cite{CristianJoan2020,CristianJoan2022}).  Next we can perform the identifications on both sides of the subtracted equation. In particular, this renders specific expressions for the scale  shifts  $\delta\MPl^2(m,M,M_0)\equiv\MPl^2 (M)-\MPl^2 (M_0)$ and $\delta\alpha(M,M_0)\equiv\alpha(M)-\alpha(M_0)$, which we need not quote here\cite{CristianJoan2020, CristianJoan2022} (see, however, our Sec.\,\ref{Sec:EffectiveActionl}). After performing these identifications, what is left of  the difference $\langle T_{\mu\nu}^{\rm vac}\rangle_{\rm Ren}(M)- \langle T_{\mu\nu}^{\rm vac}\rangle_{\rm Ren}(M_0)$ must be zero. This  demonstrates that the quantity $\delta\rL(m,M,M_0)\equiv\rL(M)-\rL(M_0)$ exactly cancels against the difference of the second term of \eqref{RenVDEexplicit} at the two scales\footnote{The (finite) differences, say $\delta\rL(m,M,M_0)$,  in this renormalization scheme should not be confused with the UV-divergent counterterms of the MS scheme, for example Eq.\,\eqref{deltaMSB}.   In the present case the renormalization has already been performed \textit{ab initio} from the subtraction prescription \eqref{EMTRenormalized}.}:
\begin{equation}\label{eq:deltarL}
\begin{split}
&\left.\rL(M)\right|_{M_0}^M+\left.\frac{1}{128\pi^2 }\left(-M^4+4m^2M^2-3m^4+2m^4 \ln \frac{m^2}{M^2}\right)\right|_{M_0}^M\\
&=\delta\rL(m,M,M_0)+\frac{1}{128\pi^2}\left(-M^4+M_0^{4}+4m^2(M^2-M_0^{2})-2m^4\ln  \frac{M^{2}}{M_0^2}\right)=0\,.
\end{split}
\end{equation}
In the absence of quartic mass terms after  renormalization, there is no need for fine tuning anymore.  Consequently, the relation \eqref{eq:VEDscalesMandM0Final} becomes substantiated.
Furthermore, Eq.\,\eqref{eq:EqsVac2}  implies that in Minkowski spacetime $\langle T_{00}^{\rm vac}\rangle_{\rm Ren}(M)=0$; and then via \eqref{RenVDE} entails that  the renormalized VED in flat spacetime is  zero,  $\rv^{\rm Mink}=0$.  In other words, zero physical CC in Minkowski spacetime. This is a must for any physical prescription aiming at a meaningful renormalization of the VED in cosmological spacetime! But of course, more conditions are needed, in particular the absence of fine tuning, as indicated above.

Let us now extract some interesting physical consequences from this renormalization framework. The  VED $\rv(M)$  as given by  Eq.\,\eqref{eq:VEDscalesMandM0Final} is not only a function of the scale $M$, but also of $H$ ($=\mathcal{H}/a$)  and its time derivatives, i.e. $\rv(M,H,\dot{H},...)$.  For the sake of more clarity, we extend mildly our abridged notation and  denote it as $ \rv(M,H)$ just to stand out its $H$ dependence. Let us now  take  two cosmic epochs $H$ and $H_0$ accessible to our observations (typically $H_0$ denotes the current epoch and $H$ one in our near past). We can relate the renormalized VED values at points  $M_0$ and $M$  from Eq.\,\eqref{RenVDEexplicit} upon  neglecting the ${\cal O}\left(H^4\right)$ terms and higher. We find
\begin{equation}\label{DiffHH0MM0}
\begin{split}
\rv(M,H)-\rv(M_0,H_0)&
=\frac{3\left(\xi-\frac{1}{6}\right)}{16\pi^2}\left[H^2\left(M^2-m^2+m^2\ln\frac{m^2}{M^2}\right)\right.\\
&\left.-H_0^2\left(M_0^2-m^2+m^2\ln\frac{m^2}{M_0^2}\right)\right]+\cdots\,,
\end{split}
\end{equation}
Notice that we have used the important Eq.\,\eqref{eq:deltarL}, which insures the cancellation of the quartic mass terms.
The physically measurable difference between the VED values at different epochs of the cosmic evolution within our observational range now follows from the usual RG prescription (cf. Sec.\ref{Sec:FlatSpaceAnalysis})  based on choosing the renormalization points near the corresponding values of the  energy scale, in this case $M_0=H_0$  and $M=H$ (and hence bringing  the renormalization point near the characteristic energy scale  of the FLRW spacetime at the  given epoch).  Denoting respectively by $\rv(H)$  and   $\rv(H_0)$ the values of $\rv(M=H,H)$ and $\rv(M_0=H_0,H_0)$  and recalling that the higher order powers are negligible for the late universe, we may cast the final result as follows:
\begin{equation}\label{eq:RVMcanonical}
\rv(H)\simeq \rvo+\frac{3\nueff(H)}{8\pi}\,(H^2-H_0^2)\,\mpl^2=\rvo+\frac{3\nueff(H)}{\kappa^2}\,(H^2-H_0^2)\,.
\end{equation}
\new{This equation constitutes the canonical form of the  RVM\,\cite{JSPRev2013}, but now derived from our QFT framework.
 In it,  $\rvo\equiv\rv(H=H_0)$  is identified with today's VED value\,\footnote{It should be emphasized that, in this QFT context,  $\rvo$  is \textit{not} at all a fundamental constant:  it is just the value of the VED at present, $H=H_0$. There is in fact no true cosmological constant in this framework, the vacuum energy density,  $\rv(H)$,  is always evolving  with the expansion, although very mildly $\propto\nueff H^2$.  What we call cosmological constant is nothing but $\CC=8\pi G_N\rv(H=H_0)=8\pi G_N\rvo$.  At any cosmic time  $t$ characterized by $H(t)$  there is a (different)   `CC' term $\CC(H)=8\pi G_N\rv(H)$ acting (approximately)  as a cosmological constant for a long period around that  time, but there is no true CC valid  at all times!},  and  $\kappa^2\equiv8\pi G_N$.  In actual fact, in the QFT context  a more precise form for the effective coupling emerges\,\cite{CristianJoan2022}:
\begin{equation}\label{eq:nueff2}
\nueff(H)\equiv\frac{1}{2\pi}\,\left(\xi-\frac16\right)\,\frac{m^2}{\mpl^2}\left(-1+\ln \frac{m^2}{H^{2}}-\frac{H_0^2}{H^2-H_0^2}\ln \frac{H^2}{H_0^2}\right)\,.
\end{equation}
It is apparent that  $\nueff(H)$  is approximately constant since it varies very slowly with the Hubble rate\footnote{We call it $\nueff$ rather than $\nu$ since its final value depends on the number and type of fields involved and it must eventually be fitted to observations, see Sec.\,\ref{sec:PhenoApplications}. }. In fact, the last term of \eqref{eq:nueff2} is logarithmic and becomes quickly suppressed for increasingly large values of $H$ above $H_0$, whereas the second term  furnishes (on account of  $\ln \frac{m^{2}}{H^2}\gg1$) the dominant (and virtually constant) contribution around the current universe ($H\simeq H_0$) and even for a large span around it\,\cite{CristianJoan2022}:
\begin{equation}\label{eq:nueffAprox2}
\nueff\equiv\nueff(H_0)\simeq\frac{1}{2\pi}\,\left(\xi-\frac{1}{6}\right)\,\frac{m^2}{\mpl^2}\ln\frac{m^2}{H_0^2}\,.
\end{equation}
The vanishing of $\nueff$ and hence of  the dynamical $\sim H^2$ part of the VED  is obtained only for conformal coupling: $\xi=1/6$.
Obviously, the leading masses $m$  involved here should come from the heaviest fields, presumably of a  Grand Unified Theory (GUT). Although we expect $|\nueff|\ll1$  owing to the ratio $m^2/\mpl^2$, the effective value of  $\nueff$ need not be negligible if one takes into account the large multiplicity of the states in a  typical GUT\,\cite{Fossil08}.  In addition, $\nueff$  also receives  fermionic contributions (independent of $\xi$) but  shall not be addressed  here.   In practice  $\nueff$   must be fitted to the  cosmological data, see Sec. \ref{sec:PhenoApplications}.}

\section{Renormalizing the EMT from  the effective action formalism}\label{Sec:EffectiveActionl}
An alternative derivation of the main renormalization formulas is possible using the effective action of the quantum  matter vacuum effects of QFT in curved spacetime, $W$\,\cite{BirrellDavies82,ParkerToms09,Fulling89}:
\begin{equation}\label{eq:DefW}
\langle T_{\mu\nu}\rangle=-\frac{2}{\sqrt{-g}} \,\frac{\delta W}{\delta g^{\mu\nu}}\,.
\end{equation}
It follows the pattern \eqref{eq:deltaTmunu2}, but here the functional variation of $W$  provides the VEV of the  EMT induced by the quantum matter vacuum  fluctuations.  As we saw in Sec.\,\ref{Sec:FlatSpaceAnalysis}, these quantum effects can be computed through a loopwise expansion in powers of $\hbar$. If the expansion is truncated at the one-loop level it involves all terms of the complete theory to order $\hbar$.  For the free field theory based on the action \eqref{eq:Sphi},  we have no self-interactions of $\phi$  and so the one-loop effective action reduces to the one-loop ZPE, and the latter yields the  exact result.
The effective action $W$  provides the quantum matter vacuum effects on top of the classical action and generalizes in curved spacetime the formula \eqref{eq:V1} used in Minkowskian space:
\begin{equation}\label{eq:EAW}
\begin{split}
W= &\frac{i\hbar}{2}Tr \ln (-G_F^{-1})
= -\frac{i\hbar}{2} \int d^4 x \sqrt{-g}\lim\limits_{x\to x'} \ln\left[ -G_F(x,x')\right]\equiv \int d^4 x \sqrt{-g}\, L_W\,.
\end{split}
\end{equation}
The last equality introduces the  Lagrangian density ${\cal L}=\sqrt{-g}\, L_W$,  in which the piece $L_W$ will be referred to as the (effective) quantum vacuum Lagrangian. It encodes the vacuum  effects from the quantized matter fields (i.e. from the scalar field $\phi$ in ouur case).   The action $W$ (and of course the Lagrangian $L_W$)  accounts for the  vacuum-to-vacuum (`bubble') matter diagrams,  i.e. the closed loop diagrams of $\phi$  without external tails, whereby it defines the zero-point energy  contributions (ZPE): a pure quantum effect.   Despite we restituted $\hbar$ provisionally in the above equation to emphasize its quantum origin,  we shall  henceforth  continue with $\hbar=1$.

In the previous sections, we have determined the  ZPE by direct computation of  the VEV of the enery-momentum tensor, i.e. the \textit{l.h.s.} of Eq.\,\eqref{eq:DefW}. Here we wish to dwell further on this result by first computing  $W$ and then inferring the  vacuum EMT from  Eq.\,\eqref{eq:DefW}.  While the procedure is well established \cite{BirrellDavies82,ParkerToms09,Fulling89} in the on-shell case, we need to carefully  track the off-shell effects as we  also did  in the previous sections from our definition of the subtracted  EMT off-shell.
In curved spacetime, the Feynman propagator, $G_F$, is the solution to the  differential equation
$\left(\Box_x-m^2-\xi R(x)\right)G_F(x,x^\prime)=-\left(-g(x)\right)^{-1/2}\delta^{(n)}(x-x^\prime)$,
where $\delta^{(n)}$ is the Dirac $\delta$ distribution in $n$ spacetime dimensions.  Although $n=4$ in our case, we keep it general since DR will be used in  intermediate calculations.   It is convenient to modify the  on-shell propagator equation in an appropriate way as follows:
\begin{equation}\label{KGPropagatorOffShell}
\left(\Box_x-M^2-\Delta^2-\xi R(x)\right)G_F(x,x^\prime)=-\left(-g(x)\right)^{-1/2}\delta^{(n)}(x-x^\prime)\,.
\end{equation}
The quantity
$\Delta^2\equiv m^2-M^2$
was defined  previously  (cf. Sec.\ref{sec:WKB}), but here it acquires a different perspective that may help to better assess its meaning. The strategy behind the modified propagator equation is to adiabatically expand its solution in the arbitrary off-shell regime $M$. We can recover the on-shell case $M=m$ by simply setting $\Delta=0$.  Nevertheless, since $\Delta^2$ is used as a probe to explore the off shell regime it must be considered of adiabatic order higher than $M$ (which is of order zero). Hence  $\Delta^2$ is assigned to be of adiabatic order $2$  (as in Sec.\,\ref{sec:WKB}), which is the next-to-leading adiabatic order consistent with general covariance.  Given that  the term  $\xi R$   is also of adiabatic order $2$, the combination $\Delta^2+\xi R$  can be dealt with as a block  of second adiabatic order.  Because the adiabaticity  of the terms must be respected in hierarchial order, the adiabatic expansion of the solution to Eq.\,\eqref{KGPropagatorOffShell} will generate new ($\Delta^2$-dependent) terms which must be taken into account as they provide important corrections to the standard on-shell result.

The adiabatic expansion method to solve the propagator equation in curved spacetime  is well-known\,\,\cite{BirrellDavies82,ParkerToms09} but it will be outlined here  since we have to track  the modification introduced by the presence of the terms  $\Delta^2$ on top of  the usual solution.  In this way we can pinpoint the effective Lagrangian defined in Eq.\,\eqref{eq:EAW} and formulate it as an asymptotic DeWitt-Schwinger expansion, as follows\,\cite{CristianJoan2022}:
\begin{equation}\label{eq:effLagrangian}
\begin{split}
L_W=&\frac{\mu^{4-n}}{2(4\pi)^{n/2}} \sum_{j=0}^\infty \hat{a}_j (x) \int_0^\infty (is)^{j-1-n/2}e^{-iM^2 s}ids
=\frac{\left({M}/{\mu}\right)^{\varepsilon}}{2(4\pi)^{2+\frac{\varepsilon}{2}}}\sum_{j=0}^\infty \hat{a}_j (x) M^{4-2j}\Gamma \left(j-2-\frac{\varepsilon}{2}\right)\,,
\end{split}
\end{equation}
where $\varepsilon\equiv n-4$ and $\mu$ is  't Hooft's mass unit.  The modified DeWitt-Schwinger coefficients for  zero, second and fourth adiabatic orders, including the $\Delta^2$-effects, read respectively as follows:
\begin{equation}\label{eq:ModifDWScoeff}
\begin{split}
&\hat{a}_0 (x)=1=a_0 (x),\\
&\hat{a}_1 (x)=-\left(\xi-\frac{1}{6}\right)R-\Delta^2=a_1(x)-\Delta^2 ,\\
&\hat{a}_2 (x)=\frac{1}{2}\left(\xi-\frac{1}{6}\right)^2R^2+\frac{\Delta^4}{2}+\Delta^2 R \left(\xi-\frac{1}{6}\right)-\frac{1}{3}Q^\lambda_{\ \lambda}=a_2(x)+\frac{\Delta^4}{2}+\Delta^2 R \left(\xi-\frac{1}{6}\right)\,,
\end{split}
\end{equation}
where the hatless $a_j(x)$ represent the ordinary coefficients when $\Delta=0$ (on-shell expansion). The term  $Q^\lambda_{\ \lambda}$ is the trace of a HD  tensor which has no impact for the FLRW spacetimes -- see \cite{CristianJoan2022} for details -- and therefore we shall not carry it along any further.

Following a procedure similar to our definition of adiabatically renormalized EMT, see Eq.\,\eqref{EMTRenormalized}, we  define now the renormalized  quantum vacuum Lagrangian  at the scale $M$. It is obtained by subtracting the first three nonvanishing adiabatic orders (the UV-divergent ones) from its on-shell value,  $L_W (m)$, at that scale, i.e.\footnote{\new{We use DR to help in identifying the cancelation of divergent terms. This is  only for convenience, as the  subtraction \eqref{eq:LWrenormalized} insures already  a finite result. Still, the intermediate UV-terms can be cancelled quite efficiently with DR.  If desired, one can obtain the same results with the subtraction procedure used to renormalize  the EMT.  We illustrated already this fact in the Appedix B of \cite{CristianJoan2020}. The important point here is not about regularization but about renormalization:  indeed, the subtraction prescription \eqref{eq:LWrenormalized} involves not just the UV-divergences  but the full expression obtained from the sum of the first three terms ($j=0,1,2$)  in the DeWitt-Schwinger expansion \eqref{eq:effLagrangian} . This is, of course,  entirely different from the MS renormalization prescription. } }
\begin{equation}\label{eq:LWrenormalized}
L_W^{\rm ren}(M )= L_W (m)-L_W^{(0-4)th}(M)\equiv  L_W (m)-L_{\rm div} (M)\,,
\end{equation}
where  $L_{\rm div}$  is the divergent part  of $L_W$, i.e. that part of  \eqref{eq:effLagrangian} involving only  the terms  up to fourth adiabatic order ($j=0,1,2$): $L_{\rm div}\equiv  L_W^{(0-4)th}$.
Upon expanding  Euler's  $\Gamma$-function in the limit $\varepsilon\to 0$  in Eq.\,\eqref{eq:effLagrangian} and using the explicit form of the modified DeWitt-Schwinger coefficients \eqref{eq:ModifDWScoeff}, one finds (after some tedious calculations) the following result:
\begin{equation}\label{eq:LWrenM}
\begin{split}
L_W^{\rm ren}(M)=\delta \rho_\Lambda(M)-\frac{1}{2}\delta\MPl^2(M) R-\delta \alpha_2(M) R^2+\cdots\,,
\end{split}
\end{equation}
 where the dots stand for irrelevant effects coming both from the mentioned HD term $Q^\lambda_{\ \lambda}$  and from subleading contributions which decouple at large $m$.  The relevant terms are listed:
\begin{equation}\label{eq:deltacouplings}
\begin{split}
&\delta\rL(M)=\frac{1}{8\left(4\pi\right)^2}\left(M^4-4m^2M^2+3m^4-2m^4 \ln \frac{m^2}{M^2}\right),\\
&\delta\MPl^2(M) =\frac{\left(\xi-\frac{1}{6}\right)}{(4\pi)^2}\left(M^2-m^2+m^2\ln \frac{m^2}{M^2}\right),\\
&\delta{\alpha_2}(M)=\frac{\left(\xi-\frac{1}{6}\right)^2}{4(4\pi)^2}\ln\frac{m^2}{M^2}\,,
\end{split}
\end{equation}
where  $\alpha_2=\alpha/2$  (recall the coefficient $\alpha$ from Einstein's equations \eqref{eq:EqsVac2}).
No dependency  is left on the unphysical  scale  $\mu$, which has cancelled altogether along with the poles at $n=4$.
Therefore, the quantities \eqref{eq:deltacouplings} stand for finite renormalization effects (cf.  footnote on p. 20).  Their precise relation with the parameter differences at the two scales $M$ and $M_0$ introduced in Sec.\,\ref{sec:RenormEMT} is obtained by setting $M_0=m$ in them:  $\delta\rL(M)=\delta\rL(m,M,m), \, \delta\MPl^2(M)=\delta\MPl^2(m,M,m)$ and $\delta{\alpha_2}(M)=\delta{\alpha_2}(M,m)$.
The renormalized vacuum effective action built up from \eqref{eq:LWrenM}, viz.
\begin{equation}\label{eq:Wren}
W_{\rm ren}(M)\equiv\int d^4 x \sqrt{-g} \  L_{\rm W}^{\rm ren}(M)=\int d^4 x \sqrt{-g} \ \left(L_W (m)-L_{\rm div} (M)\right)\,,
\end{equation}
\begin{equation}\label{eq:effActionLren}
\begin{split}
=\int d^4 x\sqrt{-g} \left( \delta \rL(M)-\frac{1}{2}\delta\MPl^2(M) R-\delta \alpha_2(M) R^2\right)\,,
\end{split}
\end{equation}
can now be readily used to  recompute  the renormalized vacuum EMT that we have found previously in Sec.\,\ref{sec:RenormEMT}.
Indeed, bearing in mind that  $\delta\alpha=2\delta\alpha_2$  one can check that
\begin{equation}\label{eq:DefWMM0}
\begin{split}
\langle T_{\mu\nu}^{\delta \phi}\rangle_{\rm ren}(M)=&-\frac{2}{\sqrt{-g}} \,\frac{\delta W_{\rm ren}(M)}{\delta g^{\mu\nu}}\\
=& \delta \rL(M) g_{\mu\nu}+\delta\MPl^2(M) G_{\mu\nu}+\delta\alpha(M) ^{(1)}{\rm H}_{\mu \nu}\,,
 \end{split}
\end{equation}
 in which  the coefficients of the various tensor expressions on the \textit{r.h.s.}  are given explicitly  by Eqs.\,\eqref{eq:deltacouplings}.
If we  take the $00th$-component of this result and use the formulas given in the appendices of \cite{CristianJoan2020,CristianJoan2022} to perform the identifications on both sides  we encounter once more the result \eqref{Renormalized2} and, with it,  we end up anew with the renormalized VED as given in Eq.\,\eqref{RenVDEexplicit} (q.e.d.)

Finally, we can derive the RG equation for the VED.  The explicit quantum correction $\delta\rL(M)$ computed in \eqref{eq:deltacouplings} adds up to the classical term $\rL(M)$  in the original action \eqref{eq:EHm} (written with renormalized parameters), such that the combined quantity $-\rL(M)+\delta\rL(M)$ must be RG invariant. From this simple observation, the $\beta$-function coefficient for the coupling $\rL(M)$ obtains:
\begin{equation}\label{eq:betarhoLambda}
  \beta_{\rL} (M)=M \frac{\partial \rL(M)}{\partial M}=\frac{1}{2(4\pi)^2}(M^2-m^2)^2\,.
\end{equation}
Using this result, the sought-for $\beta$-function for the RG-running of the vacuum energy density immediately follows from Eq.\,\eqref{RenVDEexplicit}:
\begin{equation}\label{eq:RGEVED1}
\begin{split}
\beta_{\rv}=&M \frac{\partial \rv(M)}{\partial M}
=\left(\xi-\frac{1}{6}\right)\frac{3 {H}^2 }{8 \pi^2 }\left(M^2-m^2\right)+\left(\xi-\frac{1}{6}\right)^2 \frac{9\left(\dot{H}^2 - 2 H\ddot{H} - 6 H^2 \dot{H} \right)}{8\pi^2}
\end{split}
\end{equation}
\new{where for convenience we have presented  the final result in terms of the Hubble rate in the ordinary cosmic time.
Noteworthy, no $H^4$-term is present, and, in addition, we find that all dependency on the quartic mass scales has disappeared. Compare it with the awful  $\sim m^4$ behavior obtained in \,Sec.\,\ref{Sec:FlatSpaceAnalysis}   within the MS scheme :  $\beta_{\rL}= m^4/(32\pi^2)$.  In contrast,  thanks now to \eqref{eq:betarhoLambda}  the undesired quartic terms from the renormalization of the  $\rL$ coupling  just cancel against the quartic terms from the renormalization of the ZPE, leaving Eq.\,\eqref{eq:RGEVED1}, which is a completely smoothed final result thanks to the presence of the powers of the Hubble rate and its time derivatives.  Thus,  no trace remains of those quartic mass terms in the physical result. It follows, that the VED evolution from one scale $M$ to another $M_0$ is completely free from these troublesome contributions, as shown by Eq.\,\eqref{eq:VEDscalesMandM0Final}. The net outcome is that  no need of fine tuning is ever required in this formulation, as  the running of the VED comes out to be $\propto m^2 H^2$ at most (not $\propto m^4$!), whence  a mild  function of the Hubble rate (and its derivatives).}

The following consideration may be in order.  Notwithstanding our finding that the VED evolves smoothly with the cosmic expansion, this does not mean that we can compute the actual value of the VED, $\rvo$.   Renormalization theory uses data at one point to make predictions at other points,  but it does not aim at self-predicting the initial data, of course\,\cite{Collins84}. These data  can  be used in our approach  as  boundary conditions for the RG flow to  predict the evolution of the VED at other expansion history epochs.  By the same token, it is not possible to predict the value of $H_0$ or other quantities directly related to $\rvo$.  But once these are taken as inputs,  the entire cosmic evolution becomes available in this framework and appears to  be free from unnatural adjustments among the parameters. Finally, let us mention  that this same QFT  framework can be used to show that the EoS of the quantum vacuum is not exactly $-1$ but  $\Pv/\rv=-1+\epsilon(\dot{H})$, with a small deviation $\epsilon(\dot{H})$ which in leading order involves a time derivatives of $H$.  One finds that such a deviation (induced by quantum effects) may be non-negligible in the current universe and makes the quantum vacuum mimic quintessence\,\cite{CristianJoan2022}.  A detailed study of the quantum vacuum EoS throughout the full span of the cosmic evolution will be provided elsewhere.

\section{RVM-inflation}\label{sec:Inflation}

No  less remarkable is that the renormalization of the vacuum energy density in cosmological spacetime through the  adiabatic procedure provides also a definite prediction for a dynamical mechanism of early inflation characterized by a short period where $H$=const. Such constancy of $H$, however,  has nothing to do with the ground state value of a potential (being absent in our framework)  and hence the trigger is not based on an \textit{ad hoc} inflaton field.  Such an alternative form of  inflationary mechanism is called  RVM-inflation.  To bring about RVM-inflation with an early regime of (approximately) constant Hubble rate, we need (even) powers of $H$ beyond $\sim H^2$.  Since the power $H^4$ is not explicitly available for inflation owing to the subtraction prescription \eqref{EMTRenormalized}, the next-to-leading one appears at $6th$ adiabatic order. These terms are rather cumbersome, but finite and hence do not require renormalization.  We refrain from providing the gory details here.  Calculations show that  the $6th$-order  terms relevant for this consideration are\,\cite{CristianJoan2022}
\begin{equation}\label{eq:RVMinflation}
\rv^{\rm inf}=\frac{\langle T_{00}^{\delta \phi}\rangle^{\rm 6th}_{\rm Ren}(m)}{a^2}=\frac{\txi}{80\pi^2 m^2}\, H^6+ f(\dot{H}, \ddot{H},\dddot{H}...)\,,
\end{equation}
where the superindex `inf'' denotes inflation,  and  we have defined the parameter
\begin{equation}\label{eq:xitilde}
 \txi=\left(\xi-\frac16\right)-\frac{2}{63}-360\left(\xi-\frac16\right)^3\,.
\end{equation}
The remaining terms are collected in the function $ f(\dot{H}, \ddot{H},\dddot{H}...)$ and  carry along many different combinations of powers of $H$ accompanied  in all cases with derivatives of $H$, and hence they  all vanish for $H=$const.  It means that,  up to $6th$ adiabatic order,  the only nonvanishing terms  for $H$=const. are the isolated terms proportional to the high power  $H^6$.

During the short inflationary period, the Hubble rate evolves very little around an initial (big) value $H_I$, specifically it behaves as
$H(a)=H_I{\left[1+{\left({a}/{\astar}\right)^{8}}\right]^{-1/4}}\simeq H_I$ for  $0<a<\astar$, where  $\astar$ is the transition point from a period of vacuum dominance into another of radiation dominance.  The vacuum rapidly decays into radiation and  one may compute the explicit form of the fast evolving energy densities\,\cite{CristianJoan2022}:
\begin{eqnarray}\label{rhodensities}
\rho_r(a)=\rI\,\frac{{\left({a}/{\astar}\right)^{8}}}{\left[1+{\left({a}/{\astar}\right)^{8}}\right]^{\frac{3}{2}}}\,\,, \ \ \ \ \ \ \ \ \
\rv(a)=\frac{\rI}{\left[1+{\left({a}/{\astar}\right)^{8}}\right]^{\frac{3}{2}}}\,.
\end{eqnarray}
We can easily see that at the beginning ($a=0$) there is no radiation at all ($\rho_r(0)=0$)  while the vacuum energy is maximal  ($\rv(0)=\rI$) but finite, and hence there is no initial singularity.  For $a\gg\astar$ we retrieve the standard decaying behavior of the radiation, $\rho_r(a)\sim a^{-4}$, whilst the vacuum energy density decreases very fast.  We remark that for such a very early stage of the evolution we have assumed   $\nueff=0$ and neglected any constant term  $c_0$ which might eventually become relevant in the late universe (see Eq.\,\eqref{eq:EffLambda} below).   If these terms were maintained, the asymptotic solution would recover the canonical RVM form Eq.\,\eqref{eq:RVMcanonical} that is  characteristic of  the current universe.  It is important to note that in order to trigger inflation through a large value of $H=$const.  the scalar field mass $m$ must be in the ballpark of  a GUT scale, although we should recall here that  $\phi$  has nothing to do with the  inflaton since $\phi$ has no effective potential.  Actually, we need not specify any particular property of it apart from its mass, see \,\cite{CristianJoan2022} for more details.

RVM-inflation was studied phenomenologically in a variety of works in the past, see \cite{LimaBasSol} and \cite{JSPRev2015,GRF2015,Yu2020}. These studies were based on a generalized VED of the form
\begin{equation}\label{eq:EffLambda}
\rL(H)=\frac{3}{\kappa^2}\left(c_0+\nu H^2+\alpha \frac{H^{2p+2}}{H_I^{2p}}\right)\ \ \ \ \ \ \ \  \ \ \ \  (p=1,2,3,...)\,.
\end{equation}
Here, $c_0$  is a constant of dimension $+2$ in natural units  closely related to $\Lambda$, and $H_I$ is the scale of inflation;  $\nu$ and $\alpha$ are dimensionless coefficients.   For $\alpha=0$  we recover the low-energy form of the RVM, Eq.\,\eqref{eq:RVMcanonical}, but only in the presence of higher powers of $H$ beyond $H^2$  inflation becomes possible.  In fact,  in the early universe the term $H^{2p+2}$ is dominant for  $p\geqslant 1$ and one can show that it can drive inflation. Once more the participation of only even powers of $H$  is a requirement of general covariance of the effective action.  The QFT-based model of inflation \eqref{eq:RVMinflation} just corresponds to $p=2$, for which $H^6$ is the driving force of inflation.
We should nevertheless  point out that although the mechanism of inflation based on the higher power of the VED \eqref{eq:EffLambda} was amply studied phenomenologically in the aforementioned references, it had never been substantiated on fundamental grounds.  The  demonstration that  RVM-inflation is a consequence  of calculations within QFT  in curved spacetime  has been accomplished only very recently in\,\cite{CristianJoan2022}, where the specific VED form \eqref{eq:RVMinflation} for inflation was explicitly derived. \new{Worth noticing is also the existence of a string-inspired version of RVM-inflation (see\,\cite{EPJ-ST,EPJ-Plus}, and particualry also  \cite{NickPhilTransac} in this Theme Issue), in which  inflation is triggered by gravitational anomalies through a mechanism of condensation of primordial gravitational waves  in the very early universe, which is capable of engendering $\sim H^4$ contributions to the VED.  It is amusing to think of  a double trigger for RVM-inflation from $H^4$ as well as from $H^6$ terms. The two mechanisms might even act in succession from the deepest levels of the very early universe, namely from the stringy epoch to the more conventional (QFT-like) one. For more details and spin-off possibilities, see \cite{BMS,NickMG16,NickTorsion}.  We note that the  $H^4$ power can also be motivated on more phenomenological grounds  from an analogy with the Casimir effect, see e.g. \cite{Carneiro2006}. Finally, we point out that despite the formal subtraction procedure in our QFT approach primarily leads to explicit $H^6$ contributions only, secondary  $H^4$ terms may still be generated from the scale setting $M=H$.  However, these additional considerations  will be discussed elsewhere.}

\section{Some phenomenological applications for the current universe}\label{sec:PhenoApplications}

After a good deal of technically difficult efforts aimed at putting the foundations of the RVM on a sound theoretical basis within  QFT in curved spacetime,  it should  now be welcome to bring forward some phenomenological applications. We may start from  the low-energy expression \eqref{eq:RVMcanonical}, or even from the generalized one containing  an additional term $\sim \dot{H}$ and a new coefficient $\tilde{\nu}$\,,which is also possible\cite{CristianJoan2020,CristianJoan2022}:
\begin{equation}\label{eq:RVMvacuumdadensity}
\rv(H) = \frac{3}{\kappa^2}\left(c_{0} + \nu{H^2+\tilde{\nu}\dot{H}}\right)+{\cal O}(H^4)\,.
\end{equation}
The additive parameter  $c_0$ is the same as in \eqref{eq:EffLambda}, fixed by $\rho_{\rm vac}(H_0)=\rvo$.   \new{Mind that  $c_0$ cannot vanish as otherwise the model would be excluded due to the absence of  an inflexion point from deceleration into acceleration in the cosmic evolution, see \cite{RVMphenoOlder,Polarski2012}. In particular, this fact alone excludes the simpler model $\rv(H)\propto H^2$ as a consistent vacuum model.  Within the class of the RVMs, however,  the condition $c_0\neq 0$ is always guaranteed.  Let us just note that an additive constant is always generated from integrating the renormalization group equation associated to the $\beta$-function \eqref{eq:RGEVED1}. }

The RVM  has  many phenomenological applications, see the aforementioned papers and also the monographs\,\cite{AdriaPhD2017,JavierPhD2021} and references therein.  Even though the small parameters $|\nu,\tilde{\nu}|\ll1$ can be fundamentally linked to  QFT  quantum effects (which is of course the touchstone for a solid theoretical underpinning of the RVM), it should be clear that they are  ultimately to be fitted to observations since the number of fields contributing can be large and we ignore the details of the underlying GUT ultimately responsible for such a dynamics. We can simplify our phenomenological analysis with the choice $\tilde{\nu}=\nu/2$.   In this way, Eq.\,\eqref{eq:RVMvacuumdadensity} takes on the suggestive  form
\begin{equation}\label{RRVM}
\rv(H) =\frac{3}{8\pi{G_N}}\left(c_0 + \frac{\nu}{12} {R}\right)\equiv \rv({R})\,,
\end{equation}
in which  ${R} = 12H^2 + 6\dot{H}$ is the curvature scalar. For this reason we may call this particular implementation  of the VED the  `RRVM'.  This scenario will suffice to illustrate the phenomenological performance of the RVMs.
Let us note that it is automatically well-behaved in  the radiation dominated era since, in it,  $R\simeq 0$  (or,  more precisely,   $R/H^2\ll 1$) and hence no conflict with the BBN is possible.  This condition is actually fulfilled also for sufficiently small values of $\nu,\tilde{\nu}$ in the general case\,\cite{ApJL2016}.  Two subclasses  of RRVM scenarios will be explored: i) type-I scenario, in which the vacuum interacts with matter, and ii)  type-II scenario, where  matter is conserved at the expense of a mildly evolving gravitational coupling $G (H)$ (which compensates for the vacuum evolution, and in this fashion can preserve the Bianchi identity).   For type-I models, the vacuum is assumed to exchange energy with cold dark matter (CDM) only:
\begin{equation}\label{eq:LocalConsLaw}
\dot{\rho}_{dm} + 3H\rho_{dm} = -\dot{\rho}_{\rm vac}\,.
\end{equation}
Solving the model provides the precise evolution of the matter densities\cite{EPL2021}:
\begin{equation}\label{eq:MassDensities}
\rho_m(a) = \rho^0_m{a^{-3\xi}}\,, \ \ \ \rho_{dm}(a) = \rho^{0}_m{a^{-3\xi}}  - \rho^0_b{a^{-3}} \,.
\end{equation}
Here $\rho_m=\rho_{dm}+\rho_b$ is the total matter density (CDM plus baryons).  We have defined  the parameter $\xi \equiv \frac{1 -\nu}{1 - \frac{3}{4}\nu}\simeq 1-\nu/4+{\cal O}\left(\nu^2\right)$.
The above formulae recover the $\CC$CDM form for  $\nu=0$ ($\xi=1$ ).  The departure of $\nu$ from zero, however,   is what makes possible a mild dynamical evolution of the vacuum:
\begin{align}  \label{Vacdensity}
\rv(a) &= \rvo + \left(\frac{1}{\xi} -1\right)\rho^0_m\left(a^{-3\xi} -1\right)\simeq \rvo + \frac14\,\nu\,\,\rho^0_m\left(a^{-3\xi} -1\right)+{\cal O}\left(\nu^2\right) \,.
\end{align}
The vacuum  behaves  as in the $\CC$CDM only for $\nu=0$, as then $\rv = \rvo $  remains constant.   For type-I models we make allowance for the possibility that the trigger of the DE effects from the  vacuum dynamics  occurred only  relatively recently (see  also \cite{Martinelli2019}, where a similar assumption is made).  For instance, one may conjecture that the vacuum became dynamical in the form \eqref{Vacdensity} only starting at some threshold redshift $z_{*}$ beyond which such dynamics  virtually disappears. \new{ The existence of a threshold $z_{*}$ of this sort is assumed here,  but it will be substantiated rigorously elsewhere within the QFT approach presented in the previous sections\,\cite{CristianJoan2022b}. } For definiteness, we take  $z_{*}\simeq  1$, as in \cite{EPL2021}.  We shall compare this option with the situation with no such threshold.  As for type II models,  matter is conserved, but the vacuum can still evolve as long as the effective gravitational coupling also evolves (even if very mildly) with the expansion, $G_{\rm eff}=G_{\rm eff}(H)$, e.g. starting from an initial value entering our fit. For the type-II model we do not assume any threshold effect as it proves to be smaller.  The approximate behavior of the type-II VED around the present time is (recall that $|\nu|\ll1$):
\begin{equation}\label{eq:VDEm}
\rv(a)=\frac{3c_0}{\kappa^2}(1+4\nu)+\nu\rho_m^{0}a^{-3}+\mathcal{O}(\nu^2)\,.
\end{equation}
Once more, for $\nu=0$ the VED remains constant, $\rv={3c_0}/\kappa^2$, and then $c_0=\CC/3$, but otherwise it shows a moderate dynamics as in the type-I case.   One can also show that the effective gravitational coupling evolves approximately as  $G_{\rm eff}(a)\simeq G_N (1+\epsilon\ln\,a)$ in the current epoch (with $0<\epsilon\ll 1$ of order $\nu$), thus confirming the very mild (logarithmic) evolution of $G$.  Because $\epsilon>0$ the gravitational strength exhibits asymptotically free behavior (i.e.  $G_{\rm eff}(H)$  becomes smaller in the past, where $H$ is larger).


\begin{figure}[!t]
\begin{center}
\includegraphics[width=6in]{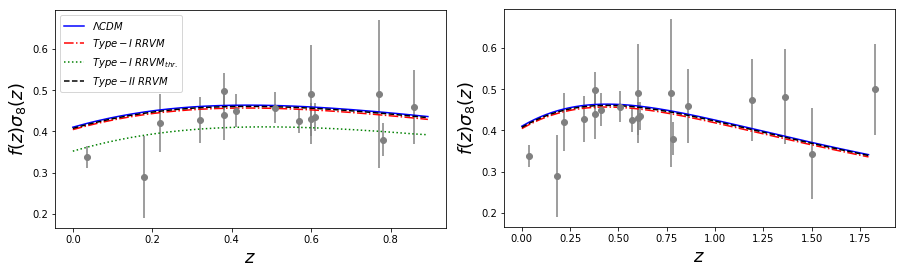}
\end{center}
{\bf Figure  1}. {\it Left plot:} Theoretical prediction for the LSS observable $f(z)\sigma_8$  corresponding to models  $\Lambda$CDM and type-I RRVM (with and without threshold redshift $z_{*}$) and type-II RRVM , confronted to the data points employed in the analysis of \cite{EPL2021}; {\it Right plot:} The same as before but in an extended redshift window.  Type-I is seen to be very successful in taming the $\sigma_8$-tension, although with little impact on the $H_0$-tension (compare with Fig.\, 2). See also Ref.\,\cite{EPL2021} for more details.
\protect\label{fig:sugra}
\end{figure}

Comparison of theoretical predictions and  observations  requires LSS  data as well, all the more if we take into account that the $\sigma_8$-tension stems from it.  Allowing for some evolution of the vacuum can be the clue to alleviate such a tension\,\cite{Intertwined8,PerivoSkara2021,TensionsJSP2018,WhitePaper2022},  as  vacuum dynamics affects nontrivially the cosmological perturbations\cite{RVMsigma8}.
We consider the perturbed, spatially flat,  FLRW metric   $ds^2=-dt^2+(\delta_{ij}+h_{ij})dx^idx^j$, in which $h_{ij}$ stands for the metric fluctuations. These fluctuations are coupled to the  matter density perturbations $\delta_m=\delta\rho_m/\rho_m$.
In our analysis \,\cite{EPL2021}, perturbations are dealt with the Einstein-Boltzmann code \texttt{CLASS}\cite{CLASS}  (in the synchronous gauge) and the statistical part with Montepython\,\cite{Montepython}.   Despite baryons are unaffected,  the corresponding equation for CDM is modified in the following way:
\begin{equation}\label{eq:perturbCDM}
\dot{\delta}_{dm}+\frac{\dot{h}}{2}-\frac{\dot{\rho}_{\rm vac}}{\rho_{dm}}\delta_{dm}=0\,,
\end{equation}
with $h=h_{ii}$ denoting the trace of $h_{ij}$. Since  the term $\dot{\rho}_{\rm vac}$ is nonvanishing for these models,  it affects the fluctuations of CDM and produces a departure from the $\CC$CDM.  The solution  must be found  numerically, as the  above perturbations equation is coupled with the metric fluctuations.
The analysis of the linear LSS regime is performed with the help of the weighted linear growth $f(z)\sigma_8(z)$, where $f(z)$ is the growth factor and $\sigma_8(z)$ is the rms mass fluctuation amplitude on scales of $R_8=8\,h^{-1}$ Mpc at redshift $z$.   The quantity $\sigma_8(z)$ is directly provided by \texttt{CLASS}. Similarly,  we can extract the  (observationally measured) linear growth function  $f(a)$ directly from the matter power spectrum $P_m(a,\vec{k})$, which is computed numerically by \texttt{CLASS}.  The results of our analysis can be seen in Figures 1 and 2 for the various models under consideration.

To compare the  RRVM's (types I and II) with the $\CC$CDM, we have defined a joint likelihood function ${\cal L}$. The total $\chi^2$ to be minimized in our case is given by\,\cite{EPL2021}
\be
\chi^2_{\rm tot}=\chi^2_{\rm SNIa}+\chi^2_{\rm BAO}+\chi^2_{ H}+\chi^2_{\rm f\sigma_8}+\chi^2_{\rm CMB}\,.
\ee
The above $\chi^2$ terms are defined in the standard way from the data including the covariance matrices\cite{DEBook}.  For the datasets used and the corresponding references, see\,\cite{EPL2021}.   In particular, the $\chi^2_{H}$ part may contain or not the local $H_0$ value measured by Riess et al.,\cite{Riess2019}. The local determination of $H_0$ (which is around $4\sigma$ away from the corresponding Planck 2018 value based on the CMB) is the origin of the so-called $H_0$ tension\cite{PerivoSkara2021,ValentinoReviewTensions,TensionsJSP2018,WhitePaper2022}.
Taking into account that the type I and II RRVM's  have one and two more parameters, respectively, as compared to the $\CC$CDM, a fairer  model comparison is achieved by computing the numerical differences between the Deviance Information Criterion (DIC) of the $\CC$CDM model against the RRVM's: $\Delta{\rm DIC}={\rm DIC}_{\rm \CC CDM}-{\rm DIC}_{\rm RRVM}$.  These differences will be (and in fact are) positive if the RRVM's fit better the overall data than the $\CC$CDM. The DIC is defined by
$  {\rm DIC}=\chi^2(\overline{\theta})+2p_D$\,\cite{InfCriteria}.
Here $p_D=\overline{\chi^2}-\chi^2(\overline{\theta})$ is the effective number of parameters of the model, and $\overline{\chi^2}$ and $\overline{\theta}$ the mean of the overall $\chi^2$ distribution and the parameters, respectively.


\begin{figure}
\begin{center}
\includegraphics[scale=0.7]{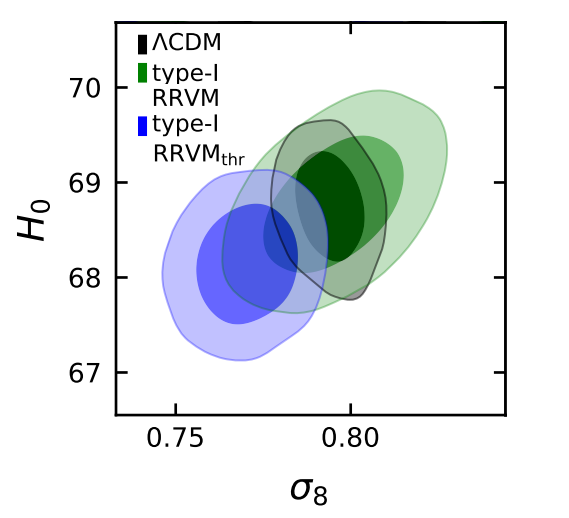}\
\includegraphics[scale=0.7]{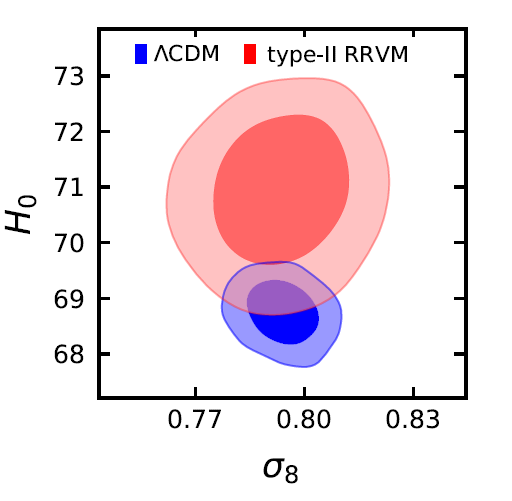}
\end{center}
{\bf Figure  2}. {\it Left plot:} $1\sigma$ and $2\sigma$ c.l. regions in the ($\sigma_8$-$H_0$)-plane. $H_0$ is expressed in km/s/Mpc units.  It is apparent that the type I RRVM with threshold redshift $z_{*}\simeq 1$  has a most  favorable impact on solving the $\sigma_8$ tension (compare with Fig.\,1); {\it Right plot:}  The corresponding regions for the type-II RRVM. One can see that the latter  is  able to significantly  curb the $H_0$ tension without worsening the $\sigma_8$ one.   $H_0$ is expressed in km/s/Mpc units.
\label{Fig:Starobinsky}
\end{figure}


The value of DIC can be computed directly from the Markov chains of our analysis\, \cite{EPL2021}.
For values $+5<\Delta{\rm DIC} <+10$ we would conclude strong support of the RRVM's as compared to the $\CC$CDM, and for $\Delta{\rm DIC}>+10$ the support becomes very strong. Such is the case when we use a threshold  redshift $z_{*}\simeq 1$ in type I RRVM.  In contrast, when the threshold is removed we find only moderate evidence against it ($-3<\Delta{\rm DIC} <-2$).  Apparently, the effect of the threshold can be very important and suggests that a mild dynamics of the vacuum is welcome, especially if it is activated at around the very epoch when the vacuum dominance appears, namely at around $z\simeq 1$ (cf. Figs. 1 and 2 left).  Unfortunately, type I models with fixed $G_{\rm eff}=G_N$ do not help an inch to solve the $H_0$ tension since the value of $H_0$ remains stuck around the CMB value\cite{EPL2021}.  In stark contrast, type II models can alleviate the two tensions at a time (cf. Fig. 2 right). The overall $\Delta{\rm DIC}$ value of the fit is quite significant ($+5.5$), still in the strong evidence region, providing values of $H_0$ markedly higher as compared to type I models (specifically $H_0=70.93^{+0.93}_{-0.87}$ Km/s/Mpc\cite{EPL2021}) along with $\sigma_8$ and $S_8$ values in the needed moderate range ($\sigma_8=0.794^{+0.013}_{-0.012}$ and $S_8=0.761^{+0.018}_{-0.017}$)\cite{EPL2021}. The values of $S_8$ in all RRVM's are perfectly compatible with recent weak lensing and galaxy clustering measurements\cite{Heymans2020}. The net outcome of this analysis is that  the only model capable of alleviating the two tensions  ($H_0$ and $\sigma_8)$ is RRVM of type II, whereas the type I model can (fully) solve the $\sigma_8$ tension but has no bearing on the $H_0$ one.  A more extensive confrontation of the RVM against the most recent cosmological data  will be presented elsewhere\,\cite{CosmoTeam2021}.  Finally, it is interesting to mention that at the  pure cosmographical  level (hence in a more model-independent way)   the RVM models \eqref{eq:RVMvacuumdadensity} appear to fit better also this kind of data than other DE models, including the concordance $\CC$CDM\,\cite{Mehdi2022a}.


\section{Conclusions and outlook} \label{sec:conclusions}

We have revisited the cosmological constant problem (CCP) and reconsidered the calculation of the renormalized energy-momentum tensor (EMT) of a real quantum scalar field non-minimally coupled to the FLRW background. Even though tackling the CCP in its full glory may require the sophisticated theoretical methods underlying quantum gravity and string theory,  difficulties appear  in all fronts:  quantum gravity is still not a well defined theory and string theory has its own frustrations.  In the meantime, we expect that some sort of provisional result  should perhaps be possible within the --  much more pedestrian -- semiclassical QFT approach, in which quantum matter fields interact with an external gravitational field, namely the FLRW background.  In this context, the renormalization of the energy-momentum tensor (EMT) of a real quantum scalar field non-minimally coupled to gravity leads to  the running vacuum model (RVM) structure, which is  the effective form of the renormalized vacuum energy density (VED). The method is based on an off-shell extension of the adiabatic regularization and  renormalization procedure, which we have used in the recent works \cite{CristianJoan2020,CristianJoan2022}.  Remarkably, it is free from  the fine tuning trouble, which is perhaps the most striking and  bizarre aspects of the  cosmological constant problem\cite{Weinberg89}. The absence of  fine-tuning in our result  is related to  the non appearance of the terms which are proportional to the quartic mass power of the field,  i.e.  $\sim m^4$,  when we relate two renormalization points.  The calculational procedure is based on the WKB expansion of the field modes in the  FLRW spacetime and  the use of an appropriately renormalized EMT. The latter is obtained by performing a substraction  from its on-shell value  (i.e. the value defined  on the mass shell  $m$ of the quantized field), specifically  we subtract the EMT computed at an arbitrary scale $M$ but only up to adiabatic order four, cf.  Eq.\,\eqref{EMTRenormalized}.  The resulting EMT becomes finite because we subtract  the first four adiabatic orders (the only ones that can be divergent in $n=4$ spacetime dimensions).  Since the off-shell renormalized EMT  becomes a function of  the arbitrary renormalization point $M$, we may compare the renormalized result at different epochs of the cosmic history, as follows.  On the one hand, the  introduction of such `sliding' scale $M$  leads to the renormalization group (RG)  flow, as in  any renormalization procedure in QFT.  On the other hand, in the case of FLRW cosmology we may associate  $M$   with Hubble's expansion rate $H$;  that is to say, we first leave $M$ arbitrary  but  set $M=H$ only at the end.  This association ties the formal RG running with the cosmological evolution and  we can determine the VED at the expansion history time corresponding to the value $H$ of the Hubble rate.  Indeed, we find that this is the consistent association in order to make contact with physics and at the same time to obtain a well-defined running of the gravitational coupling, see \cite{CristianJoan2022} for  details.

Although the RVM was originally motivated from semi-qualitative RG ideas (see\,\cite{JSPRev2013} and references therein), the current  QFT calculations in curved spacetime go far beyond them and provide for the first time a solid foundation of the  RVM, in which the dynamical structure of the VED is seen to emerge at the quantum level from the adiabatic  renormalization of the EMT\cite{CristianJoan2020,CristianJoan2022}. Thus, no special action is needed for the RVM; the structure of the VED emerges from quantum effects within the  Einstein-Hilbert action complemented with the usual higher derivative vacuum terms in curved spacetime\cite{BirrellDavies82}, including of course the quantized matter fields (in general with nominimal interactions in the scalar sector). It is to be particularly noted the interesting  implications for the very early universe. A new mechanism for inflation is predicted, called RVM-inflation\,\cite{CristianJoan2022}.  It is based on an early and very short period in which  $H=$const.  Such a constant is totally unrelated to the ground state value of a scalar field, so RVM-inflation has nothing to do with \textit{ad hoc}  inflaton fields.  The driving force of RVM-inflation is the predicted $\sim H^6$ power emerging from quantum effects on the effective action of QFT in FLRW spacetime.    The RVM inflationary mechanism is also completely distinct from Starobinsky's inflation\,\cite{Starobinsky80}, for which it is the time derivative of the Hubble rate which remains constant for a short period of time ($\dot{H}=$ const.)\cite{JSPRev2015}.    We have also pointed out the existence of  a stringy version of the RVM inflationary mechanism which can operate with  $H^4$ terms\,\cite{EPJ-ST,EPJ-Plus,NickPhilTransac}.  These terms can also appear in QFT when the setting $M=H$ is performed and will be investigated elsewhere.

 At the end of the day, we have been able to show that the genuine form of the VED for the current universe can be attained  from direct calculations of QFT in the FLRW spacetime.   In such a structure, the powers of $H$ (and its time derivatives)  are of even adiabatic order.  This means that all of the permitted powers  effectively carry an even number of time derivatives of the scale factor, which is essential to preserve the general covariance of the action.  Linear terms (cubic, or in general of odd order) in $H$ are incompatible with covariance and in fact do not appear in the final result.  To be more precise, all terms with an odd number of time derivatives of the scale factor) are ruled out by covariance and never enter our calculation.

 On the other hand, the form of  $\rho_{vac}(H)$ predicted by the RVM  at low energies is no less interesting. It is remarkably simple but it certainly goes beyond a rigid cosmological constant term.  To be precise,  it consists of a dominant additive term  (to be identified basically with the cosmological term)  together with a small dynamical component  $\sim \nu H^2$, in which the dimensionless parameter $\nu$ can be computed from the underlying QFT framework and is predicted to be small  ($|\nu|\ll1$).  Ultimately, its value can only be known  upon fitting the RVM to the overall cosmological data, where it has been found to be positive and in the ballpark of $\sim 10^{-3}$\, cf.\cite{EPL2021}.  Such  a mild dynamics of the cosmic vacuum is  helpful  to ameliorate the global fits as compared to the concordance  $\CC$CDM. The $H_0$ tension can be reduced  to $\sim 1.6\sigma$  and the $\sigma_8$  one at $\sim 1.3\sigma$.  These significantly lower levels of tension are perfectly tolerable and should cause no concern. The successful cutback of the two tensions (for type-II RRVM's)  is highly remarkable and is strongly supported by standard information criteria, such as the deviance information criterion (DIC). \new{Let us note that the type I and II  versions of the RVMs,  which we have investigated phenomenologically here, represent an approximation to the full  RVM structure derived in the previous sections.  Such an approximation was necessary in order to test more easily the phenomenological impact.  However, remarkably enough, it can be shown that the full RVM derived from QFT actually mimics simultaneously the two model types that we have studied here,  and therefore it benefits from the phenomenological  virtues of them both\,\cite{CristianJoan2022}.}

 Obviously, much more work and the analysis of a wealth of upcoming data in the next few years will be needed to confirm the phenomenological soundness of the RVM\,\cite{CosmoTeam2021}.   However, irrespective of the fate of the aforementioned phenomenological tensions in the future to come,  two  important theoretical facts will remain: i)  There is no true cosmological constant $\CC$  in the RVM framework;  QFT in FLRW spacetime replaces it with  a dynamical term, $\CC(H)$, which undergoes a very mild cosmic evolution $\sim\nueff  H^2$  (except in the early universe, where higher powers of $H$  triggered inflation);  and  ii) thanks to a more physical  renormalization  of the vacuum  energy density within QFT in curved spacetime,  no fine tuning is needed to insure  a smooth dynamical evolution of the VED throughout the cosmic epochs.

 We close with the following observation. In spite of the fact that  our study focused on the zero-point energy of a free scalar field non-minimally coupled to gravity, we conjecture that similar results should hold good in the general case with interactions, since the expansion of the effective action in powers of momenta in the context of FLRW spacetime should essentially result in a momentum-independent value (the VEV of the effective potential) plus an expansion in (even) powers of $H$ (thanks to general covariance).

\vspace{1.5cm}

{\bf Acknowledgements}: {The author thanks Prof. S. Cotsakis and Prof. Alexander Yefremov  for the invitation to contribute to this Special Theme Issue of the Phil. Trans. of the Royal Society A,  ``The Future of Mathematical Cosmology''. This work is funded in part by projects  PID2019-105614GB-C21,  FPA2016-76005-C2-1-P (MINECO, Spain), 2017-SGR-929 (Generalitat de Catalunya) and CEX2019-000918-M (ICCUB, Barcelona). The author also acknowledges participation in the COST Association Action CA18108 ``{\it Quantum Gravity Phenomenology in the Multimessenger Approach (QG-MM)}''. Last, but not least, I am   grateful  to A.  G\' omez-Valent,  J. de Cruz P\'erez and C. Moreno-Pulido  for  the fruitful collaboration in the task of understanding the RVM and its manyfold phenomenological implications.}



\end{document}